\documentclass[preprint,12pt]{elsarticle}

\usepackage{amsthm,amsmath,amsfonts,amssymb}
\usepackage{mathtools}
\usepackage{makecell}
\usepackage{color}
\usepackage{bm}
\usepackage{algorithm}
\usepackage{algorithmicx}
\usepackage{algpseudocode}
\usepackage{graphicx}
\usepackage[caption=false]{subfig}
\usepackage{threeparttable}
\usepackage{multirow}
\usepackage[colorlinks,citecolor=blue,urlcolor=blue]{hyperref}

\theoremstyle{plain}
\newtheorem{theorem}{Theorem}[section]

\newtheorem{assumption}{Assumption}
\newtheorem{proposition}[theorem]{Proposition}

\begin{document}
	
	\begin{frontmatter}
		
		\title{Double Descent, Ensemble Emergence, and Large Model Averaging in High-Dimensional Multimodel Prediction}
		
		\author[1]{Ke Chen}
		\ead{kechen@stu.xjtu.edu.cn}
		\author[1]{Dandan Jiang}
		\ead{jiangdd@xjtu.edu.cn}
		\author[2]{Xinyu Zhang\corref{cor1}}
		\ead{xinyu@amss.ac.cn}
		
		\affiliation[1]{organization={School of Mathematics and Statistics, Xi'an Jiaotong University}}
		\affiliation[2]{organization={Academy of Mathematics and Systems Science, Chinese Academy of Sciences}}
		
		\cortext[cor1]{Corresponding author.}

\begin{abstract}
This paper investigates the predictive performance of high-dimensional multimodel prediction, where the number of regressors is comparable to the sample size. 
Leveraging tools from random matrix theory, we derive the exact limiting out-of-sample risk under a nested model setting and comprehensively characterize the risk landscape.
This limiting risk helps to reveal two phenomena:
simple weighting inherits the double descent trajectory and its associated variance explosion near the interpolation boundary;
strategic weighting triggers an ensemble emergence that suppresses the localized risk surge and yields a globally flat risk surface.
Building on this limiting risk, we also propose the Large Model Averaging (LaMA) method, in which we consider the discrepancy between in-sample and out-of-sample risks in the high-dimensional regime. 
Numerical studies and real data applications confirm that LaMA achieves superior predictive accuracy in high-dimensional environments.
\end{abstract}


\end{frontmatter}



\section{Introduction}
The fundamental objective of modern statistical learning is to achieve robust prediction through out-of-sample generalization. 
This pursuit encounters a structural challenge in high-dimensional regimes where the number of regressors $p$ grows proportionally with the sample size $n$.
For the individual candidate models that serve as the building blocks of an ensemble, the true out-of-sample predictive risk deviates from the classical U-shaped paradigm. 
Instead of increasing monotonically after passing the complexity threshold, the risk exhibits a double descent trajectory (\cite{doi:10.1073/pnas.1903070116,doi:10.1137/20M1336072}). 
As rigorously proven by \cite{10.1214/21-AOS2133} for single linear models, the out-of-sample risk diverges sharply near the interpolation boundary ($p=n$) due to a variance explosion, before descending again in the over-parameterized regime. 
Unlike model selection, which relies on a single chosen candidate, model averaging incorporates all available information by combining diverse candidate models (\cite{10.1257jel}). 
Therefore, a natural idea is to alleviate this extreme risk in a single model through ensembling.

The predictive performance of high-dimensional multimodel prediction depends on its weight allocation.
Equal or random weights reduce the ensemble risk by averaging candidate models of different complexities, but such indiscriminate weights cause the estimator to inherit the estimation variance of its components (\cite{NEURIPS2020_7d420e2b,HDMA2023}).
Thus, the risk surface preserves the double-descent shape, and this risk peak near the interpolation boundary may impair generalization and numerical stability.
Our study also reveals that by strategically allocating smaller weights to high-risk candidates, the model averaging framework not only suppresses the variance explosion that arises from near-interpolating candidate models, but also transforms the divergent double-descent ridge of a single model into a globally flat risk surface, a phenomenon we characterize as ensemble emergence.
However, this transformation of the out-of-sample risk landscape remains overlooked in the existing model averaging literature.

When the model dimension scales proportionally with the sample size, the sample covariance matrix ceases to be a consistent estimator of the population covariance matrix in the spectral norm (\cite{10.1016/S0047, Bai2010}).
The inconsistency in the spectral properties leads to a discrepancy between the behavior of in-sample risk and that of true out-of-sample predictive risk.
Traditional frequentist methods, such as those based on the Mallows model averaging (MMA) criterion (\cite{MMA2007,WAN2010277,Liu25112016}) or cross-validation (\cite{HANSEN201238}), rely on in-sample risk to track predictive performance.
This approximation breaks down severely near the interpolation boundary, rendering these criteria ineffective where this discrepancy is highly pronounced. 
Concurrently, existing high-dimensional approaches (\cite{Ando2014AMA,Ando2017MA,https://doi.org/10.1002/sta4.317,31e34903-9988-3190-86d8-6e5af63558c4,10.1093/jrsssb/qkae094}) 
typically rely on feature screening or marginal correlation grouping to restrict the maximum dimension  of candidate model $k_M$ much smaller than the sample size $n$.
This is a common practice to maintain theoretical tractability, but such a strategy does not directly mitigate the potential risk explosion when the dimension grows proportionally with $n$. 
Although \cite{Zhang02042020,Zou2025} allow diverging dimensions, their stringent rate constraints (e.g., $k_M=o(n^{1/4})$ in \cite{Zhang02042020} and $k_M=o(n^{1/2})$ in \cite{Zou2025}) do not allow for the situation where $k_M/n$ tends to a constant.
Moreover, their asymptotic optimality theories still target in-sample risk, which limits their applicability in high-dimensional multimodel prediction.

To address this misalignment between in-sample and out-of-sample risks, we propose a new model averaging method, Large Model Averaging (LaMA). 
First, LaMA adjusts the traditional MMA criterion, which targets the unbiased estimation of in-sample risk, by substituting the limiting out-of-sample variance for its in-sample counterpart under a high-dimensional setting.
Second, to encourage a smoother weight distribution that incorporates more candidate information and adaptively penalize candidates with high estimation variance, we introduce a variance-weighted $\ell_2$ penalty on the weight vector.
Through these corrections, the design criterion naturally decomposes into the estimation of in-sample bias, the asymptotic out-of-sample variances, and an $\ell_2$ regularization term. 
This construction yields a dual-risk balancing regularization that considers both fitting accuracy and generalization ability. 
Consequently, rather than ignoring the generalization gap between in-sample and out-of-sample risks, LaMA accounts for this discrepancy in the optimization objective, thereby improving robustness against the ``curse of high dimensionality''.

Our main contributions are threefold.
Theoretically, using limit spectral theory for high-dimensional random matrices, we derive the limiting behavior of the out-of-sample risk in model averaging under nested models.
Through the three-dimensional map of the limiting risk surface, we identify the double-descent phenomenon under simple weighting and the ensemble emergence with strategic weight allocation.
Methodologically, for high-dimensional multimodel prediction, the proposed LaMA moves beyond the traditional reliance on in-sample-risk-based alternative criteria, achieving statistically grounded weight calibration without artificial dimension reduction under high-dimensional regimes.
Practically, our method does not require the classical assumption that the model dimension must be much smaller than the sample size,
and remains effective even when the two are comparable.
By avoiding the unstable risk peak near the interpolation boundary, LaMA achieves competitive predictive performance, especially when the dimension is comparable to the sample size.

\section{Model formulation}\label{sec: Model}
Based on the work of \cite{MMA2007}, we consider the data-generating process 
$y_i = \mu_i + e_i, \ i=1,\cdots, n,$
where \(\mu_i\) is modeled as a linear combination of an infinite series of explanatory variables and its form is 
$\mu_{i} = \sum_{j=1}^{\infty} \theta_{j} x_{ij}$.
Let \(\bm{x}_{i}=(x_{i1}, x_{i2},\cdots)^{\prime} \) is countably infinite and
\(\theta_{j}\) is the corresponding coefficient.
Furthermore, $y_i$ is a real-valued sample observed from the response variable, and 
\(e_{i}\) is the independent random error with \(\mathbb{E} \left[e_{i}|\bm{x}_{i}\right] = 0\) and \(\mathbb{E}\left[e_{i}^{2}|\bm{x}_{i}\right] = \sigma^{2}\).
It is also assumed that $0 < \sigma^2 < \infty$ and $\mathbb{E}\left[\mu_i^2\right] < \infty$ for all $i$.

The choice of different combinations of explanatory variables corresponds to the formulation of distinct approximate predictive models.
Consider a series of linear approximation models 
\begin{equation}\label{model1}
	y_i = \sum_{j=1}^{k_q} \theta_{j(q)} x_{ij(q)} +b_{i(q)}+ e_i, \quad q=1,\cdots, M,
\end{equation}
where the $q$th model utilizes \(k_{q}(>0)\) regressors, \(\theta_{j(q)}\) are the corresponding coefficients, and \(b_{i(q)} = \mu_{i} - \sum_{j=1}^{k_{q}} \theta_{j(q)} x_{ij(q)}\) is the approximation error. 
The subset \(\{x_{ij(q)}, j=1,\cdots, k_q\}\) is  selected from the comprehensive set \(\bm{x}_{i}\in \mathbb{R}^p (p>k_q)\) with zero mean and population covariance matrix \(\bm{\Sigma}\).
To express the above model in matrix form, the variables and parameters are represented as matrices, defined as follows:
\(\bm{Y} = (y_{1}, \cdots, y_{n})^{\prime}\), \(\bm{\mu} = (\mu_{1}, \cdots, \mu_{n})^{\prime}\),
\(\bm{\Theta}_{q} = (\theta_{1(q)}, \cdots, \theta_{k_{q}(q)})^{\prime}\), \(\bm{b}_{q} = (b_{1(q)}, \cdots, b_{n(q)})^{\prime}\), \(\bm{e} = (e_{1}, \cdots, e_{n})^{\prime}\), and \(\bm{x}_{i(q)} = (x_{i1(q)}, \cdots, x_{ik_{q}(q)})^{\prime}\). 
The design matrix of the $q$th candidate model is  \(\bm{X}_{(q)} = (\bm{x}_{1(q)}, \cdots, \bm{x}_{n(q)})^{\prime}\), which is an \(n \times k_{q}\) matrix.
Thus, for the $q$th candidate model, the model \eqref{model1} can be written as
\begin{equation}	\label{model2}
	\bm{Y}=\bm{\mu}+\bm{e}=\bm{X}_{(q)}\bm{\Theta}_{q} +\bm{b}_q+ \bm{e}, \quad q=1,\cdots, M.
\end{equation}

For each \(q\)th  candidate  model \eqref{model2}, the minimum \(\ell_2\) norm least squares regression estimator of \(\bm{\Theta}_{q}\) is considered and defined as below
\begin{equation*}
	\hat{\bm{\Theta}}_{q} = \mathop{\mathop{\arg\min}}_{\bm{\Theta}_q \in \mathbb{R}^{k_q}}   \ \|\bm{Y} - \bm{X}_{(q)}\bm{\Theta}_q\|^2 .
\end{equation*}
Or, equivalently, it is written as 
\begin{equation}\label{MME}
	\hat{\bm{\Theta}}_{q}= (\bm{X}_{(q)}^{\prime} \bm{X}_{(q)})^+ \bm{X}_{(q)}^{\prime}  \bm{Y},   	
\end{equation}
where \((\bm{X}_{(q)}^{\prime} \bm{X}_{(q)})^+\) is the Moore-Penrose pseudo-inverse of \(\bm{X}_{(q)}^{\prime} \bm{X}_{(q)}\) to cope with the case of \(k_q>n\).
When $\bm{X}_{(q)}$ has full column rank, the estimator simplifies to the classical least squares estimator \(\hat{\bm{\Theta}}_{q} = (\bm{X}_{(q)}^{\prime} \bm{X}_{(q)})^{-1} \bm{X}_{(q)}^{\prime} \bm{Y}\). It follows that the corresponding estimator of \(\bm{\mu}\) is  expressed as  \(\hat{\bm{\mu}}_{q} = \bm{X}_{(q)} \hat{\bm{\Theta}}_{q} \),  and 
the residual vector as \(\hat{\bm{e}}_{q} = \bm{Y} - \hat{\bm{\mu}}_{q}\).
Furthermore, 
the model averaging estimators for $\bm{\Theta}$ and \(\bm{\mu}\) are given by
$$\hat{\bm{\Theta}}(\bm{\omega})=\sum_{q=1}^{M} \omega_{q} \left(\begin{array}{c}
	\hat{\bm{\Theta}}_{q}\\
	\bm{0}_{ }
\end{array}\right) \  \text{and} \ \ \hat{\bm{\mu}}(\bm{\omega})=\sum_{q=1}^{M} \omega_{q} \hat{\bm{\mu}}_{q}=\bm{P}(\bm{\omega})\bm{Y},$$
where $\bm{P}(\bm{\omega}) \coloneqq \sum_{q=1}^{M} \omega_{q}\bm{X}_{(q)} (\bm{X}_{(q)}^{\prime} \bm{X}_{(q)})^+ \bm{X}_{(q)}^{\prime}$ and \(\bm{\omega} = (\omega_{1}, \cdots, \omega_{M})^{\prime}\) is a weight vector in \(
\mathcal{H}_{n} = \{ \bm{\omega} \in [0,1]^{M} : \sum_{q=1}^{M} \omega_{q} = 1 \} \). 
In addition, $\bm{0}_{ }$ is a zero vector of appropriate dimension, ensuring that $\hat{\bm{\Theta}}(\bm{\omega}) \in \mathbb{R}^p$.

The primary goal of model averaging is to improve predictive accuracy by minimizing the out-of-sample loss:
$$L_{\text{out}}(\bm{\omega}) = \left(\hat{\mu}_0(\bm{\omega}) - \mu_0\right)^2,$$
where $\hat{\mu}_0(\bm{\omega}) = \sum_{q=1}^M \omega_q \bm{x}_{0(q)}^\prime \hat{\bm{\Theta}}_q$ is the prediction for a new observation $\bm{x}_0$, and $\mu_0 = \bm{x}_0^\prime \bm{\Theta}$ is the true conditional mean, given an independent test sample $\{y_0, \bm{x}_0\}$.
Traditionally, model averaging methods operate under the implicit premise that the in-sample loss and out-of-sample loss are asymptotically equivalent,  
thus targeting the minimization of the in-sample loss:
$$L_{\text{in}}(\bm{\omega}) =  \|\hat{\bm{\mu}}(\bm{\omega}) - \bm{\mu}\|^2/n.$$
For instance, seminal works \cite{MMA2007, HANSEN201238} and subsequent important developments (e.g., \cite{Li_article2018, RePEc785-798, Fang_Yuan_Tian_2023}) formulate their weight choice criteria by constructing unbiased estimators of in-sample risk $\mathbb{E}\left[L_{\text{in}}(\bm{\omega})\mid \bm{X}\right]$, establishing their asymptotic optimality based on this objective.

\begin{figure*}[tbp] 
	\centering
	\subfloat[]{\includegraphics[width=0.5\linewidth]{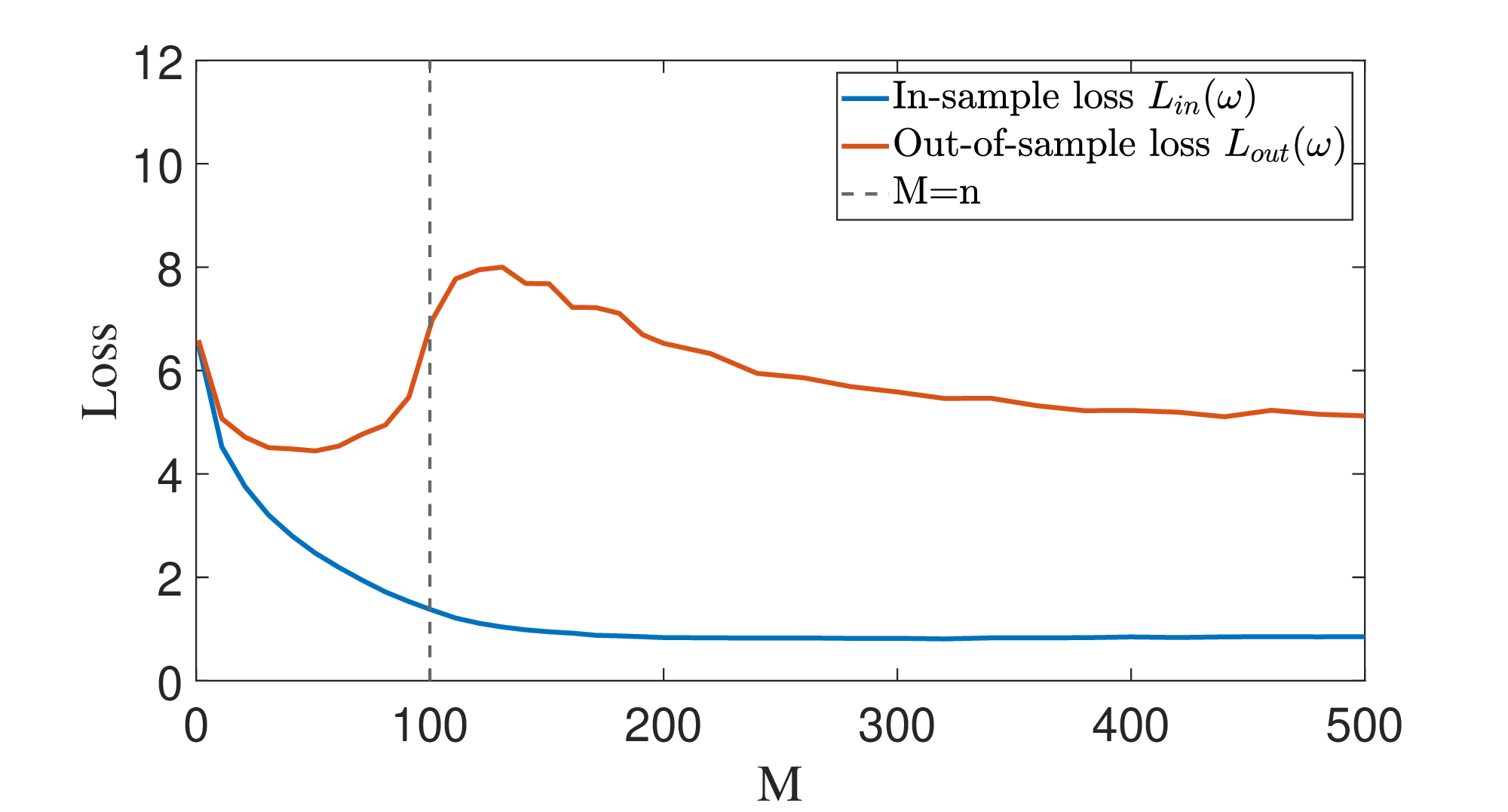}}
	\subfloat[]{\includegraphics[width=0.5\linewidth]{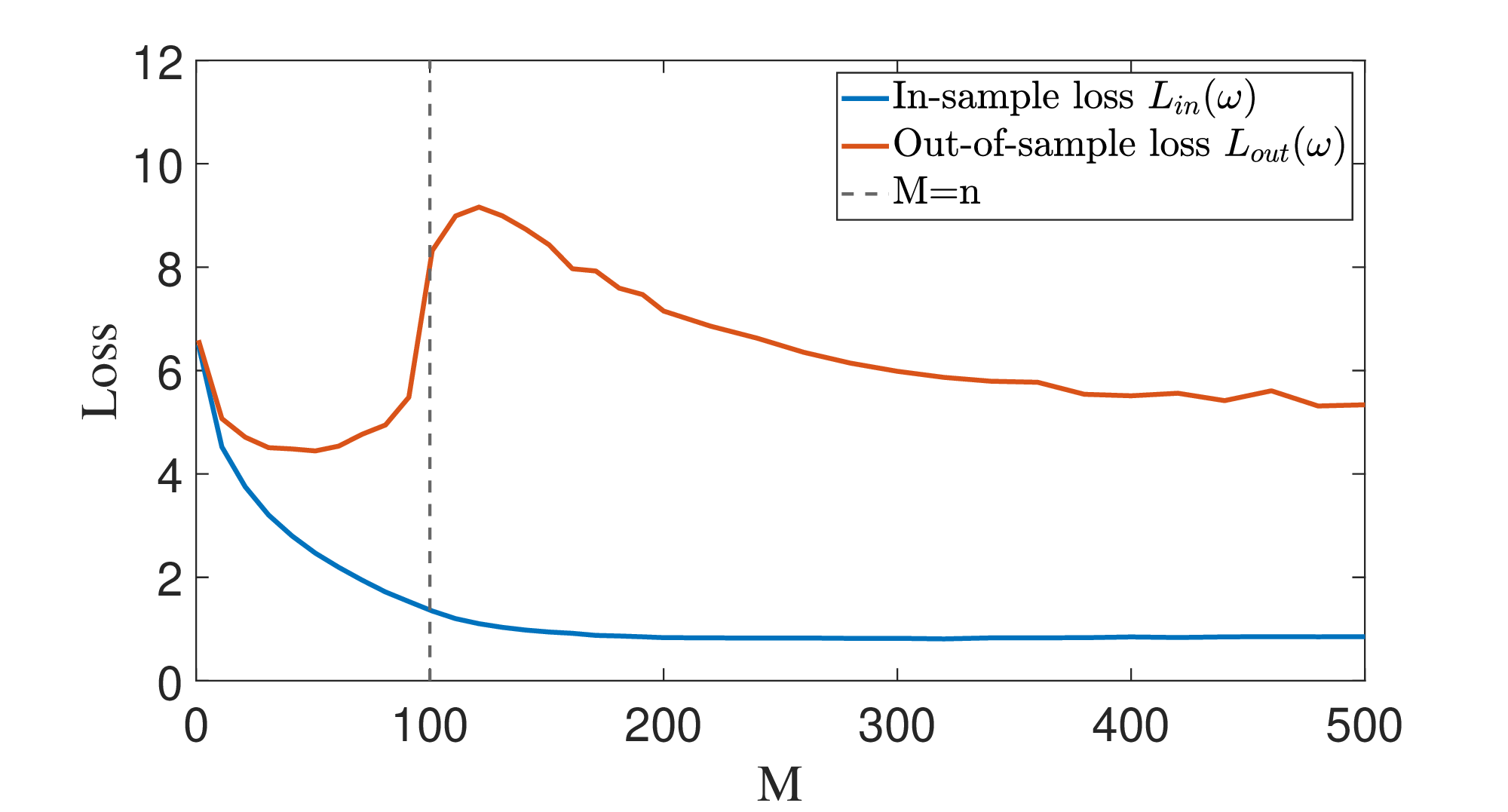}}
	\vspace{-1em}
	\caption{The trend of in-sample loss $L_{\text{in}}(\bm{\omega})$ and out-of-sample loss $L_{\text{out}}(\bm{\omega})$ as the number of candidate models $M$ increases. To address the numerical ill-conditioning at $k_q = n$ that causes loss divergence, we implement two strategies to manage the singularities at the interpolation boundary: (a) excluding the model with $k_q=n$; (b) truncating extreme loss values exceeding 50. The regression coefficient $\bm{\Theta}$ is set to $\theta_j=j^{-0.5}, j=1,\cdots,1000$. Equal weights $\omega_q=1/M $ are used for nested models $\{k_q=q\}_{q=1}^M$.}
	\label{fig: overfit}
\end{figure*}

Indeed, we observe that $L_{\text{in}}(\bm{\omega})$ closely approximates $L_{\text{out}}(\bm{\omega})$ in low-dimensional scenarios ($k_q \ll n$). However, in high-dimensional regimes where the number of regressors $k_q$ is comparable to the sample size $n$, this approximation breaks down. 
Minimizing the in-sample loss no longer guarantees optimal out-of-sample predictive accuracy. 
The details are shown in Figure \ref{fig: overfit}, which illustrates this divergence by comparing $L_{\text{in}}(\bm{\omega})$ and $L_{\text{out}}(\bm{\omega})$ as the number of candidate models $M$ increases. 
Following the convention in \cite{10.1214/21-AOS2133}, 
we refer to $M=n$ as the ``interpolation boundary'', where the largest candidate model interpolates the training data ($\|\hat{\bm{\mu}}_M - \bm{Y}\|^2/n = 0$). 
As the number of candidate models $M$ exceeds the interpolation boundary, the out-of-sample loss rises to a peak before eventually decreasing, contrasting starkly with the monotonic decline of in-sample loss. 
This misalignment becomes pronounced in high-dimensional settings, rendering traditional model averaging methods suboptimal for predictive tasks.

To rectify this, it is essential to accurately account for the out-of-sample risk during the weight choice process. 
By analytically characterizing its high-dimensional asymptotic behavior, we correct the inaccurate estimation in traditional \( L_{\text{in}}(\bm{\omega})\)-based methods, thereby constructing an efficient model averaging method that mitigates overfitting and enhances generalization ability.

\section{Double descent and ensemble emergence in asymptotic out-of-sample risk}
In this section, we leverage random matrix theory to derive closed-form limits for the out-of-sample risk via a bias-variance decomposition. 
These theoretical limits analytically characterize the double descent and ensemble emergence phenomena; we further extend these limits to general covariance structures, thereby laying the computable foundation for our proposed LaMA method.
\subsection{Limiting theory for out-of-sample risk}\label{subsec: decomposition}
Assume that the observed training data are \((y_i, \bm{x}_{i}) \in \mathbb{R} \times \mathbb{R}^{p}, i = 1, \ldots , n\), where $p$ is the total number of regressors.  
We denote the full design matrix for each $q$th model by
\( \bm{X} = (\bm{x}_1, \cdots, \bm{x}_n)^{\prime} \), where $\bm{\Sigma}$ denotes the population covariance matrix of the observations $\bm{x}_i$. 
For test data \(\bm{x}_{0}\) independent of training data \(\bm{X}\), the corresponding  out-of-sample predictive risk in model averaging is
\begin{equation}\label{Riskout}
	R_{\text{out}}(\bm{\omega})=\mathbb{E}\left[ L_{\text{out}}(\bm{\omega}) \mid \bm{X}\right]=\mathbb{E} \left[\left(\hat{\mu}_0(\bm{\omega}) -\mu_0\right)^2 \;\middle|\; \bm{X}\right],
\end{equation}
which is the conditional expectation of out-of-sample loss.
We divide the test data \(\bm{x}_{0}\) into two parts \(\bm{x}_{0}=(\bm{x}_{0(q)}^{\prime},\bm{x}_{re(q)}^{\prime})^{\prime}\), where $\bm{x}_{0(q)}$ is the vector of explanatory variables in the $q$th candidate model and $\bm{x}_{re(q)}$ is the vector of the remaining variables. Similar partitions are applied to \(\bm{\Theta}=(\bm{\Theta}_{q}^{\prime},\bm{\Theta}_{re(q)}^{\prime})^{\prime}\),  \(\bm{X}=(\bm{X}_{(q)},\bm{X}_{re(q)})\),  and 		
$$\bm{\Sigma}=\left(
\begin{array}{c c}
	(\bm{\Sigma}_{q})_{k_q\times k_q} & (\bm{\Sigma}_{q,re(q)})_{k_q\times (p-k_q)}\\
	(\bm{\Sigma}_{re(q),q})_{(p-k_q)\times k_q}  & (\bm{\Sigma}_{re(q),re(q)})_{(p-k_q)\times (p-k_q)}
\end{array}\right).$$

Recall that $\hat{\mu}_0(\bm{\omega}) = \sum_{q=1}^M \omega_q \bm{x}_{0(q)}^\prime \hat{\bm{\Theta}}_q$ and $\mu_0 = \bm{x}_0^\prime \bm{\Theta}$ denote the model averaging prediction and the true conditional mean, respectively.
By decomposing the out-of-sample risk with respect to the training response $\bm{Y}$ and the new observation $\bm{x}_0$, we obtain the following bias-variance decomposition:
\begin{equation}\label{decompose of R}
	R_{\text{out}}(\bm{\omega}) = R_{\text{out},V}(\bm{\omega}) + R_{\text{out},B}(\bm{\omega}),
\end{equation}
where the variance and bias components are defined as
\begin{align}
	R_{\text{out},V}(\bm{\omega}) &=\mathbb{E}\left[\operatorname{Var}\left(\hat{\mu}_0(\bm{\omega})\mid \bm{X},\bm{x}_0\right)\;\middle|\; \bm{X}\right]	\label{RVq}\\
	&= \sum_{q=1}^{M}\sum_{l=1}^{M} \omega_{q}\omega_{l} \operatorname{tr}\left( \operatorname{Cov}\left(\hat{\bm{\Theta}}_{l},\hat{\bm{\Theta}}_{q} \;\middle|\; \bm{X}\right) \mathbb{E}\left[\bm{x}_{0(q)}\bm{x}_{0(l)}^\prime \;\middle|\; \bm{X}\right] \right),  \notag\\
	R_{\text{out},B}(\bm{\omega}) &=\mathbb{E}\left[\left(\mathbb{E}\left[\hat{\mu}_0(\bm{\omega})\mid \bm{X},\bm{x}_0\right]-\mu_0\right)^2\;\middle|\; \bm{X}\right] \label{RB_matrix}\\
	&= \mathbb{E}\left[ \left( \sum_{q=1}^{M} \omega_{q}\bm{x}_{0(q)}^\prime\mathbb{E}\left[\hat{\bm{\Theta}}_{q} \;\middle|\; \bm{X}\right] - \bm{x}_{0}^\prime\bm{\Theta} \right)^2 \;\middle|\; \bm{X}\right].\notag
\end{align}
In these expressions, $R_{\text{out},V}(\bm{\omega})$ represents the conditional variance of prediction $\hat{\mu}_0(\bm{\omega})$ with respect to $\bm{Y}$, averaged over the new observation $\bm{x}_0$.
It thus captures how the estimation covariance of parameters $\operatorname{Cov}(\hat{\bm{\Theta}}_{l},\hat{\bm{\Theta}}_{q} \mid \bm{X})$ propagates into the final predictive variance. 
Furthermore, it measures the extent to which the randomness in the training data inflates the ensemble prediction. For this reason, we call it the out-of-sample variance.
Correspondingly, $R_{\text{out},B}(\bm{\omega})$ represents the expected squared deviation of the conditional mean prediction $\mathbb{E}\left[\hat{\mu}_0(\bm{\omega}) \mid \bm{X},\bm{x}_0\right]$ from the true response $\mu_0$.
This term quantifies the systematic error resulting from model misspecification and the approximation bias of the candidate models, and is thus termed the out-of-sample bias. 
This decomposition structure is similar to the bias-variance decomposition commonly seen in machine learning (\cite{doi:10.1073/pnas.1903070116,zhou2021machine}), clearly revealing the intrinsic mechanism of the model averaging method in balancing the estimated variance with the model specification bias. The optimal weight choice needs to strike a balance between bias and variance to minimize the total out-of-sample risk.

To understand how this bias-variance trade-off behaves in high-dimensional regime, we now turn to an asymptotic analysis. 
We consider the large-scale asymptotic setting where \(n, k_q \to \infty\)  such that \(k_q/n \to c_q\in (0, \infty)\), and study the limiting behavior of the out-of-sample risk function under two scenarios: under-parameterization (\(c_q<1\)) and over-parameterization  (\(c_q>1\)).  
Specifically, we list the assumptions in terms of high-dimensional asymptotics.
\begin{assumption}\label{assump1}
	The vector $\bm{x}_{i(q)}$ is generated from a population with zero mean and a deterministic positive definite covariance matrix $\bm{\Sigma}_q$, satisfying $\bm{x}_{i(q)} = \bm{\Sigma}^{1/2}_q \bm{z}$. 
	Here, $\bm{z}$ is a random vector with independently and identically distributed (i.i.d.) entries having zero mean, unit variance, and a finite fourth moment. 
	Furthermore, there exists a constant $\eta > 0$ such that $\lambda_{\min}(\bm{\Sigma}_q) \geq \eta$, where $\lambda_{\min}(\bm{\Sigma}_q)$ denotes the smallest eigenvalue of $\bm{\Sigma}_q$.
\end{assumption}

\begin{assumption}\label{assump3}
	The vector $\bm{x}_{i(q)}$ consists of independent entries with zero mean and covariance matrix $\bm{I}_{k_q}$. 
	The entries $x_{ij}$ possess a finite moment of order $4+\delta$, i.e., $\mathbb{E}\left[|x_{ij}|^{4+\delta}\right] < C$ for some constants $C, \delta > 0$. 
\end{assumption}

Assumption~\ref{assump1} describes a general data generation mechanism, allowing for a general covariance structure \(\bm{\Sigma}_q\).
It imposes regularity on the covariance matrices, requiring them to be positive definite with eigenvalues uniformly bounded away from zero. 
This condition ensures non-singularity and bounded spectral norms, which is a common regularity condition in random matrix theory.
Clearly, Assumption~\ref{assump3} is a special case of Assumption~\ref{assump1} when \(\bm{\Sigma}_q = \bm{I}_{k_q}\); however, it imposes a stronger moment condition than Assumption~\ref{assump1}, requiring a finite $(4+\delta)$th moment rather than merely a finite fourth moment.
This stronger condition is typically needed to obtain sharper asymptotic results or to simplify technical proofs under the independence assumption. 
In the subsequent analysis, these assumptions will be invoked appropriately depending on the specific scenarios.
The following conclusions from the classical results of random matrix theory, visible in Chapter 6 of \cite{SERDOBOLSKII2008239} and Theorem 1 of \cite{10.1214/21-AOS2133}. 
We use $\to$ to denote the standard limit of a deterministic sequence, and $\xrightarrow{a.s.}$ for almost sure convergence.

\begin{proposition}\label{limit}
	Consider the model in \eqref{model2} and suppose that Assumption \ref{assump1} is satisfied.   
	Then as $n, k_q \to \infty$ with $k_q/n \to c_q \in (0,1)$, it holds that
	$$ \operatorname{tr}\left(\left(\bm{X}_{(q)}^\prime\bm{X}_{(q)}\right)^{-1}\bm{\Sigma}_q\right) \xrightarrow{a.s.}  \frac{c_q}{1 - c_q}. $$
	Further assume that Assumption \ref{assump3} holds. Then as $n, k_q \to \infty$ with $k_q/n \to c_q \in( 1, \infty )$, it holds that
	$$\bm{\Theta}_{q}^{\prime} \bm{X}_{(q)}^\prime\left(\bm{X}_{(q)}\bm{X}_{(q)}^\prime\right)^{-1} \bm{X}_{(q)} \bm{\Theta}_{q} \xrightarrow{a.s.} \frac{1}{c_q}\left\|\bm{\Theta}_{q}\right\|_2^2 \quad \text{and}\quad
	\operatorname{tr}\left(\left(\bm{X}_{(q)}^\prime\bm{X}_{(q)}\right)^+\right)	\xrightarrow{a.s.} \frac1{c_q-1}.$$
\end{proposition}     
These limits yield the limiting out-of-sample risk for a correctly specified single linear model (see \cite{10.1214/21-AOS2133}):
\begin{equation*}
	R_{\text{single}} \xrightarrow{a.s.}
	\begin{cases}
		\sigma^2\frac{c_q}{1 - c_q},&  c_q<1\\
		\|\bm{\Theta}_{q}\|^2\left(1-\frac{1}{c_q}\right)+\sigma^2\frac{1}{c_q-1}, &  c_q>1
	\end{cases}.
\end{equation*}
Notably, in this setting, the double-descent phenomenon only manifests as the second descent when exceeding the interpolation boundary ($c_q > 1$). 
For $c_q < 1$, the risk increases monotonically because the approximation bias is zero, concealing the first descent typically driven by regressors additions in misspecified settings.

In contrast, our candidate models \eqref{model2} are inherently misspecified, introducing approximation bias via the omitted signal $\bm{b}_q$.  
Furthermore, unlike the single-model case where the bias follows the generalized Mar\v{c}enko-Pastur theorem (\cite{RUBIO2011592}), model averaging estimator introduces complex cross-model dependencies that yield intractable interaction terms in the out-of-sample risk.
To bypass this intractability while preserving the risk landscape, we analyze the asymptotic expectation of the risk to obtain a closed-form characterization.
We begin with the isotropic case where $\bm{\Sigma}=\bm{I}_p$, and then extending to general covariance matrices in Section \ref{sec: general}.
Due to space constraints, the detailed derivations involving random matrix techniques are deferred to supplementary material. 
The deterministic limit for out-of-sample risk \eqref{decompose of R} are summarized in the following theorem.

\begin{theorem}\label{Th: out-of-sample risk}
	Consider a sequence of $M$ nested candidate models with dimensions $k_1 < k_2 < \dots < k_M$ and suppose that Assumption \ref{assump3} is satisfied. For any given weight vector $\bm{\omega} \in \mathcal{H}_n$, as $n, k_q, k_l \to \infty$ such that $k_q/n \to c_q$ and $k_l/n \to c_l$, where $c_q, c_l \in (0,1) \cup (1, \infty)$, the out-of-sample risk $R_{\text{out}}(\bm{\omega})$ exhibits the following asymptotic behavior:
	
	(i) The out-of-sample variance $R_{\text{out},V}(\bm{\omega})$ in \eqref{RVq} and the expected out-of-sample bias $\mathbb{E}\left[R_{\text{out},B}(\bm{\omega})\right]$ in \eqref{RB_matrix} converge to their respective limits:
	\begin{equation}\label{RBV limit}
		R_{\text{out},V}(\bm{\omega}) \xrightarrow{a.s.} \bm{\omega}^\prime \bm{D}_V \bm{\omega} \quad \text{and} \quad \mathbb{E}\left[R_{\text{out},B}(\bm{\omega})\right] \to \bm{\omega}^\prime \bm{D}_B \bm{\omega},
	\end{equation}
	where $\bm{D}_V$ and $\bm{D}_B$ are $M \times M$ symmetric matrices. By symmetry, assuming $c_q \leq c_l$, their $(q,l)$-th entries are respectively given by
	\begin{equation}\label{var Dv}
		\bm{D}_V(q,l)=
		\begin{cases}
			\sigma^2\frac{c_{q}}{1-c_{q}} ,&  c_q\leq c_l<1\\
			\sigma^2\frac{c_{q}}{c_l-c_{q}}, &  c_q<1<c_l \\
			\sigma^2\frac{1}{c_l-1}, &  1<c_q\leq c_l
		\end{cases}
	\end{equation}
	and
	\begin{equation*}
		\bm{D}_{B}(q,l)=
		\begin{cases}
			\frac{1}{1-c_{q}}\|\bm{\Theta}_{re(l)}\|^2, &  c_q\leq c_l<1\\
			\frac{c_l-1}{c_l-c_{q}}\left(\|\bm{\Theta}_{l}\|^2-\|\bm{\Theta}_{q}\|^2\right)+\frac{c_{l}}{c_l-c_{q}}\|\bm{\Theta}_{re(l)}\|^2, &  c_q<1<c_l \\
			\frac{c_q-1}{c_{q}}\|\bm{\Theta}_{q}\|^2+\left(\|\bm{\Theta}_{l}\|^2-\|\bm{\Theta}_{q}\|^2\right)+\frac{c_l}{c_{l}-1}\|\bm{\Theta}_{re(l)}\|^2, &  1<c_q\leq c_l
		\end{cases}.
	\end{equation*}
	(ii) Furthermore, if the entries of $\bm{X}$ are i.i.d. $\mathcal{N}(0,1)$ and the signal strength satisfies $\|\bm{\Theta}\|^2 = o(\sqrt{n / \log n})$, then the out-of-sample risk holds that
	\begin{equation}\label{limit d}
		R_{\text{out}}(\bm{\omega}) \xrightarrow{a.s.} \bm{\omega}^\prime (\bm{D}_V + \bm{D}_B) \bm{\omega}.
	\end{equation}
\end{theorem}

Theorem \ref{Th: out-of-sample risk} characterizes the limiting behavior of each component of the out-of-sample risk in high-dimensional regime, revealing how the model complexity (measured by the aspect ratios \(c_q\) and \(c_l\)) and the true parameter structure jointly determine out-of-sample bias and variance.
The elements $\bm{D}_B(q,l)$ of the limiting matrix $\bm{D}_B$ for out-of-sample bias fall into three regimes:
(i)  When all model dimensions are strictly less than the sample size $(c_q\leq c_l < 1)$, the cross-bias component originates solely from the omitted parameters $\bm{\Theta}_{re(l)}$ of the larger $l$th model, and its coefficient increases monotonically as $c_q$ grows. 
(ii) In the mixed regime $(c_q<1<c_l)$, the coefficients exhibit divergent behavior as $c_q$ and $c_l$ approach the critical threshold 1. This phenomenon stems from the inherent ill-conditioning of the estimator near the interpolation boundary, which triggers a dramatic inflation of the bias components.
(iii)	 When both model dimensions exceed the sample size $(1<c_q\leq c_l)$, the contribution of the included parameter $\bm{\Theta}_{q}$ is scaled by a compression factor $1-1/c_q$, while the contribution of the omitted parameter $\bm{\Theta}_{re(l)}$ is amplified by a factor of $1+1/(c_l-1)$. This amplification effect highlights the system's sensitivity to even minor variable omissions in the overparameterized regime.
Similarly, the out-of-sample variance term \(V_n^{(q,l)}\) exhibits a piecewise limiting form, growing sharply as the dimension-to-sample ratio approaches 1.

The signal strength condition $\|\bm{\Theta}\|^2 = o(\sqrt{n/ \log n})$ is a technical requirement for almost sure convergence of the bias component.  
Unlike the constant signal norm assumption in high-dimensional linear regression (e.g., \cite{10.1214/17-AOS1549, 10.1214/21-AOS2133}), our condition allows the signal norm $\|\bm{\Theta}\|$ to diverge slowly with $n$ at a rate nearly up to $o(n^{1/4})$.  Furthermore, if the objective is relaxed from almost sure convergence to convergence in probability, this growth condition on the signal strength can be weakened.

\subsection{Visualizing double descent and ensemble emergence}
\begin{figure*}[h] 
	\centering 
	\begin{minipage}{0.64\linewidth}
		\centering
		\subfloat[]{\includegraphics[width=\linewidth]{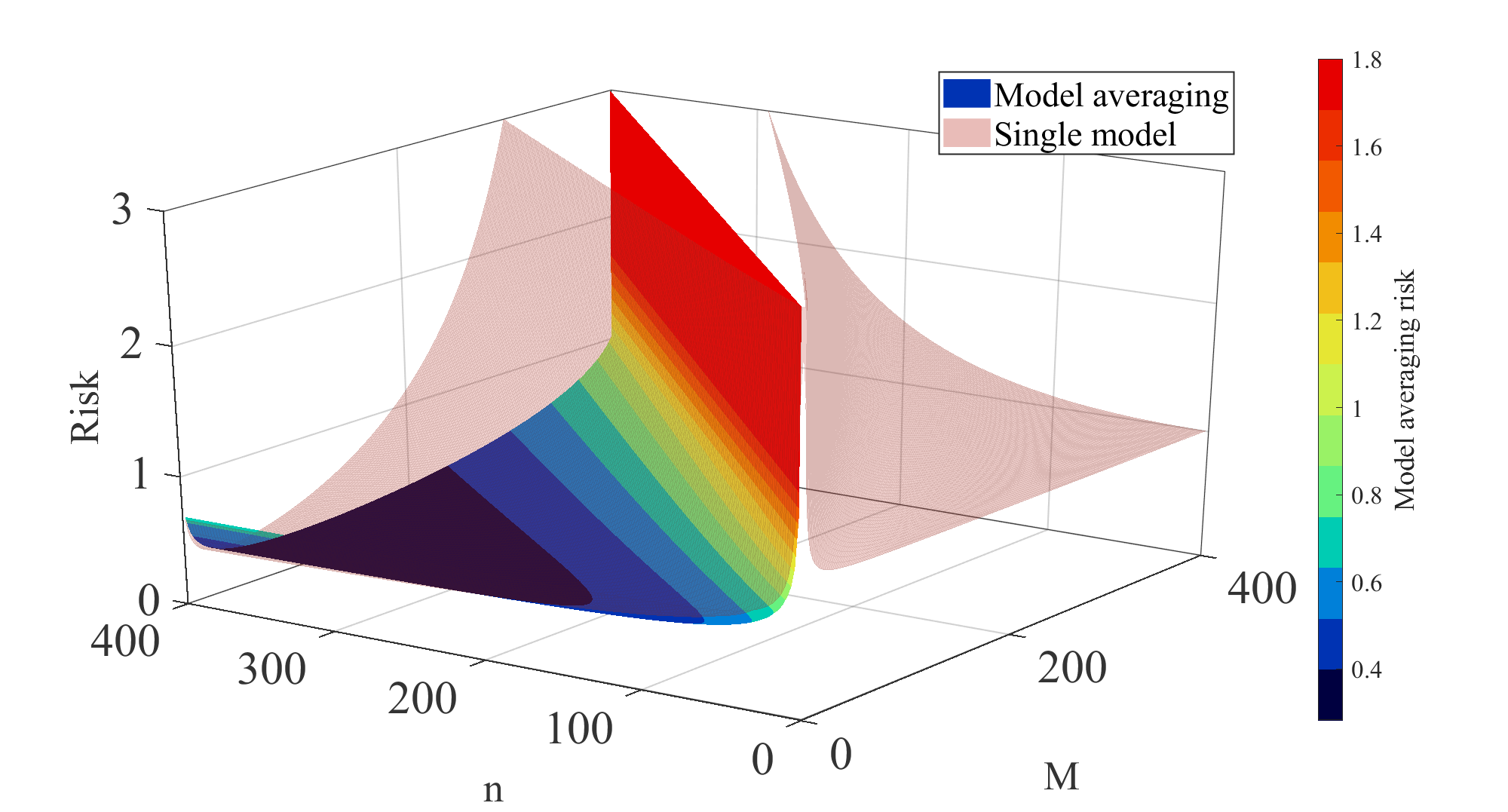}}
	\end{minipage}
	\begin{minipage}{0.35\linewidth}
		\centering
		\subfloat[fixed $n=100$]{\includegraphics[width=\linewidth]{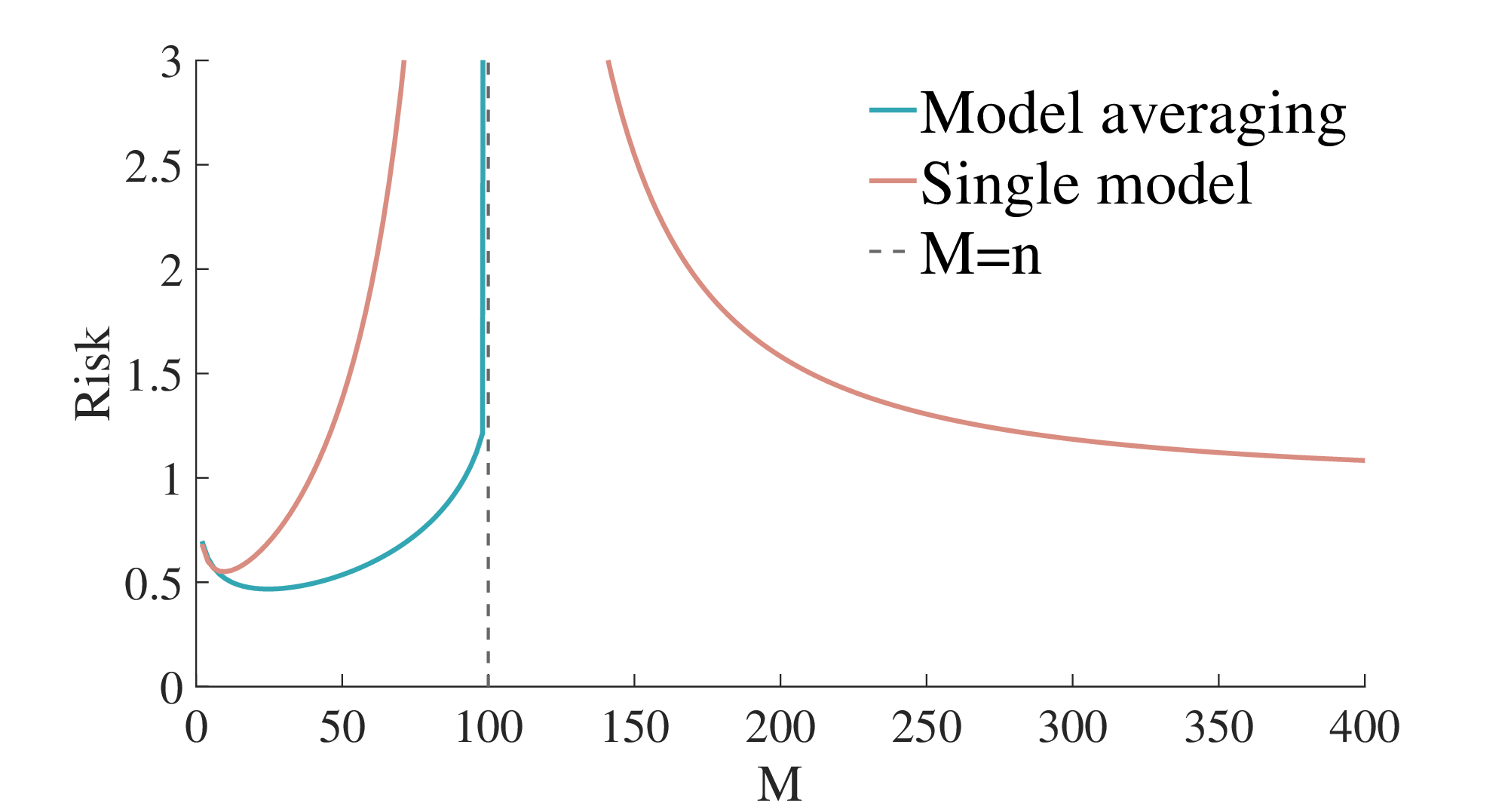}}\\
		\subfloat[fixed $M=100$]{\includegraphics[width=\linewidth]{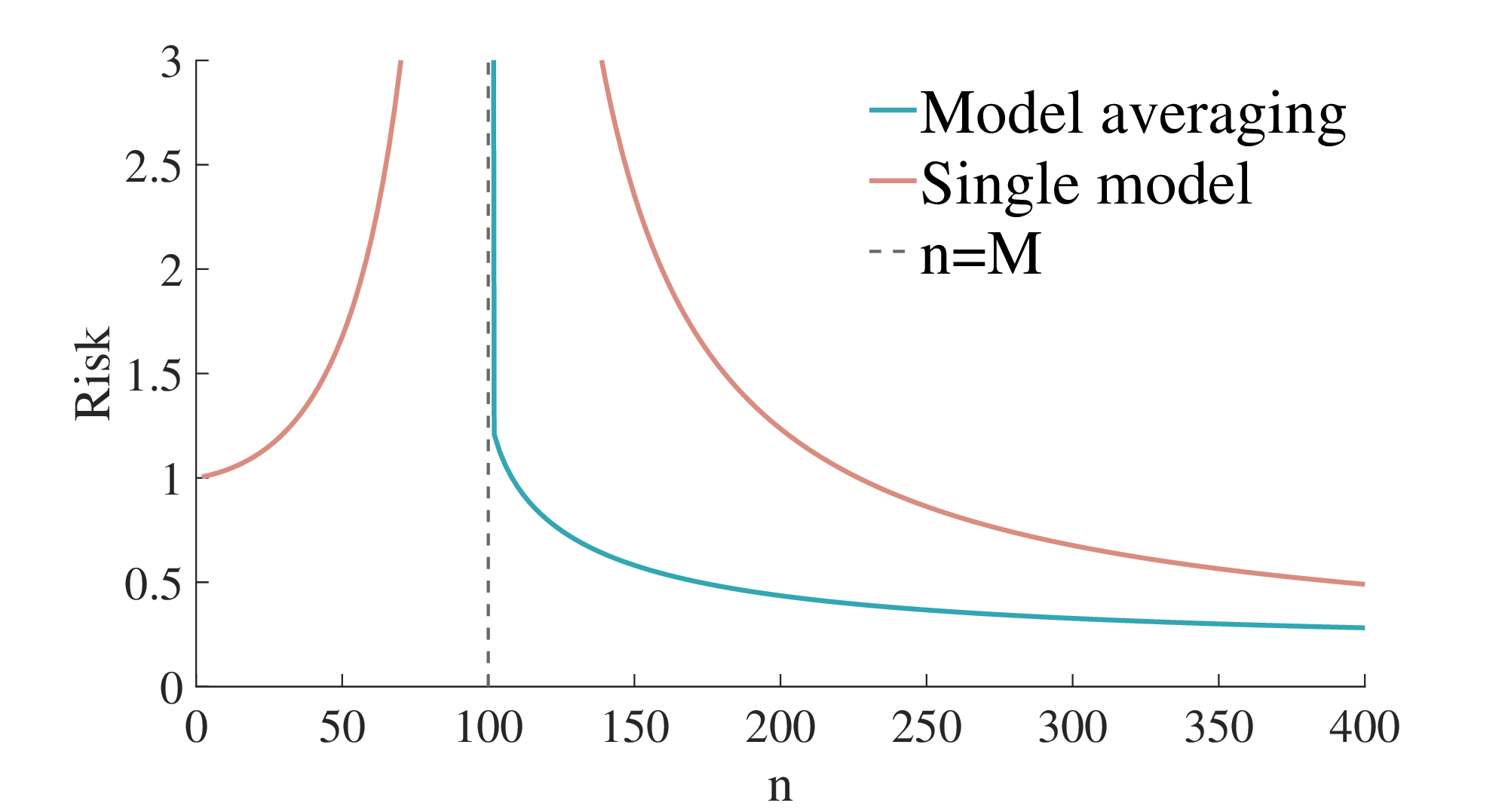}}
	\end{minipage}
	\vspace{-1em}
	\caption{Under equal weights (singular sub-models with $k_q=n$ retained), the limiting behavior of out-of-sample risk $R_{\text{out}}(\bm{\omega})$ for model averaging and $R_{\text{single}}^{(q)}$ for a  single model as $M$ and $n$ vary. 
		The signal-to-noise ratio (SNR) is  $\mathrm{SNR}=\|\bm{\Theta} \|^2/\sigma^2=1$ with $\sigma^2=1$. The regression coefficient $\bm{\Theta}$ is set to $\theta_j=\gamma j^{-0.6}, j=1,\cdots,400$, where the constant $\gamma$ is controlled by $\mathrm{SNR}$.}
	\label{fig: risk_infty}
\end{figure*}

Based on the theoretical convergence result \eqref{limit d},
we utilize the asymptotic risk surface as an approximation of the true generalization error, allowing us to reveal the intrinsic mechanisms underlying the generalization performance of model averaging.
We construct $M$ nested candidate models with dimensions $k_q=q$ for $q \in \{1, \dots, M\}$.
The ensemble risk is compared against the asymptotic risk of the largest individual candidate model, $R_{\text{single}}^{(q)}$, which by setting $M=1$ in Theorem~\ref{Th: out-of-sample risk} follows:
\begin{equation}\label{single risk}
	R_{\text{single}}^{(q)} \xrightarrow{a.s.}
	\begin{cases}
		\frac{1}{1-c_q}\|\bm{\Theta}_{re(q)}\|^2+\sigma^2\frac{c_q}{1 - c_q}   ,&  c_q<1\vspace{1ex}\\
		\frac{c_q-1}{c_{q}}\|\bm{\Theta}_{q}\|^2+\frac{c_q}{c_{q}-1}\|\bm{\Theta}_{re(q)}\|^2+\sigma^2\frac{1}{c_q-1} , &  c_q>1
	\end{cases}.
\end{equation}
As a baseline, we first examine the simple equal-weight allocation strategy.
If the candidate set includes a sub-model with $k_q=n$, its out-of-sample variance diverges as $c_q \to 1$,  causing the ensemble risk to explode for all $M \ge n$; consequently, the risk surface becomes unbounded and cannot be displayed in the $M>n$ region, as shown in Figure~\ref{fig: risk_infty}.

\subsubsection{Double descent in equally weighted model averaging}
\begin{figure*}[] 
	\centering 
	\begin{minipage}{0.64\linewidth}
		\centering
		\subfloat[]{\includegraphics[width=\linewidth]{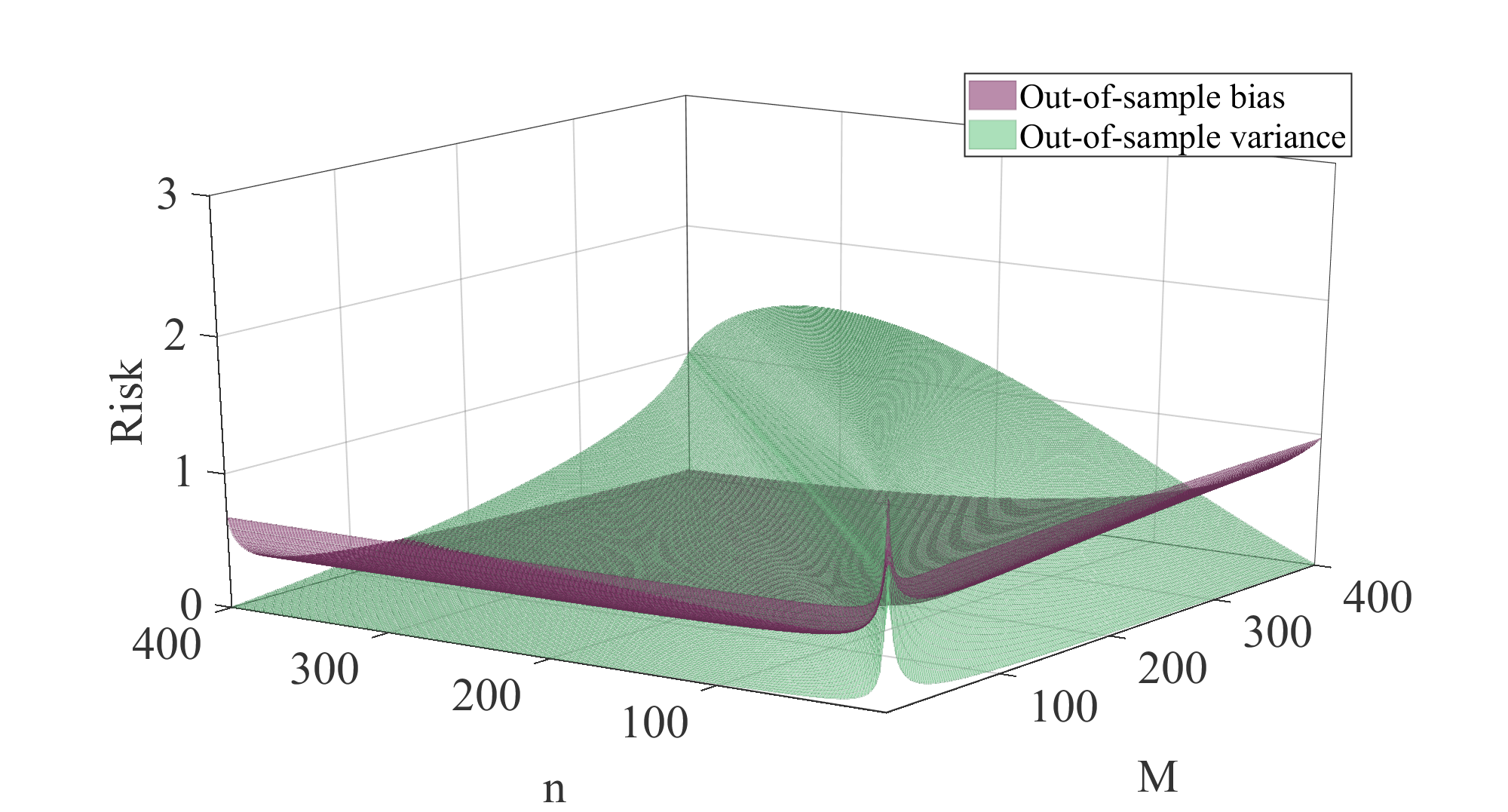}}
	\end{minipage}
	\begin{minipage}{0.35\linewidth}
		\centering
		\subfloat[fixed $n=100$]{\includegraphics[width=\linewidth]{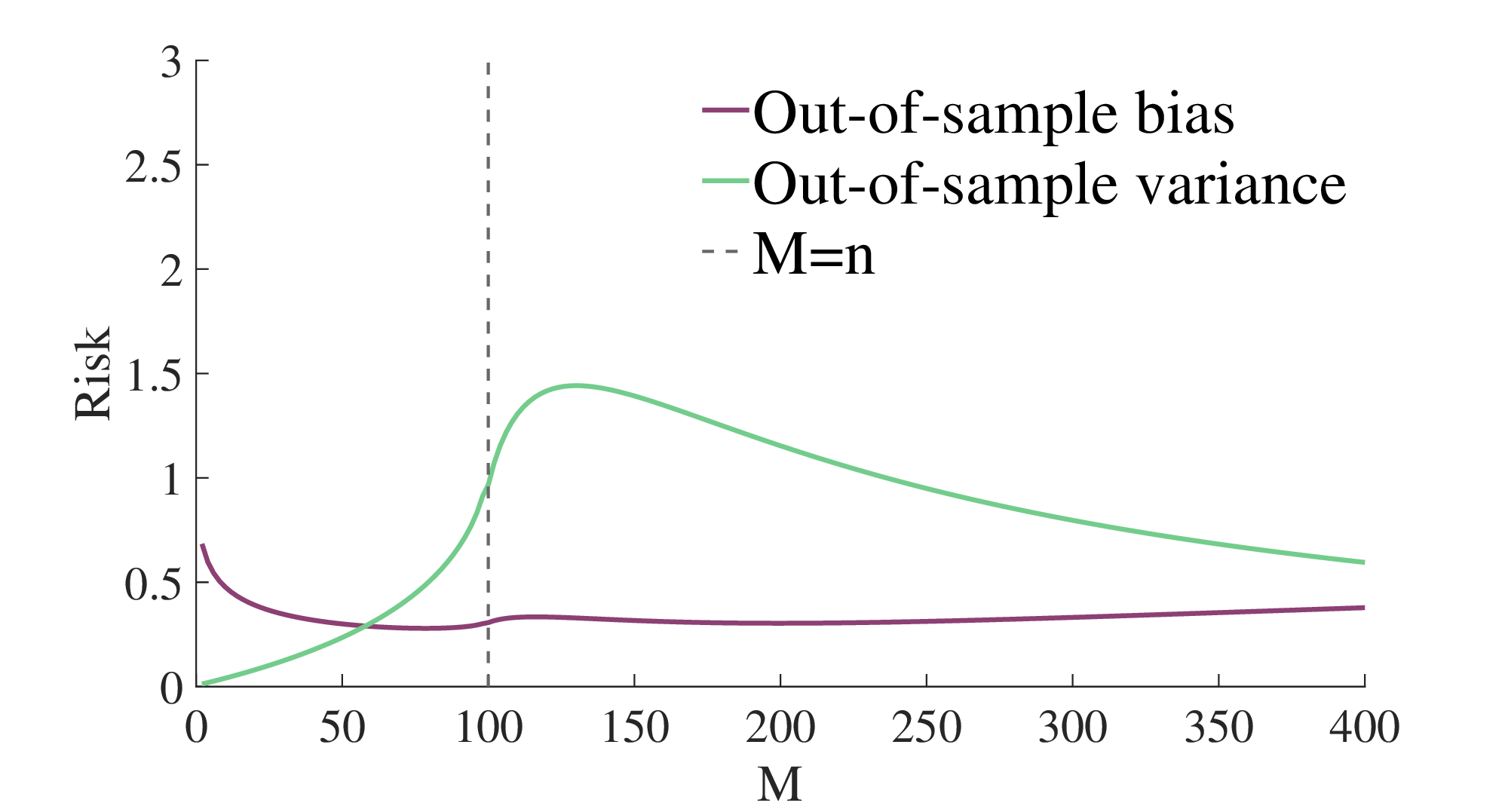}}\\
		\subfloat[fixed $M=100$]{\includegraphics[width=\linewidth]{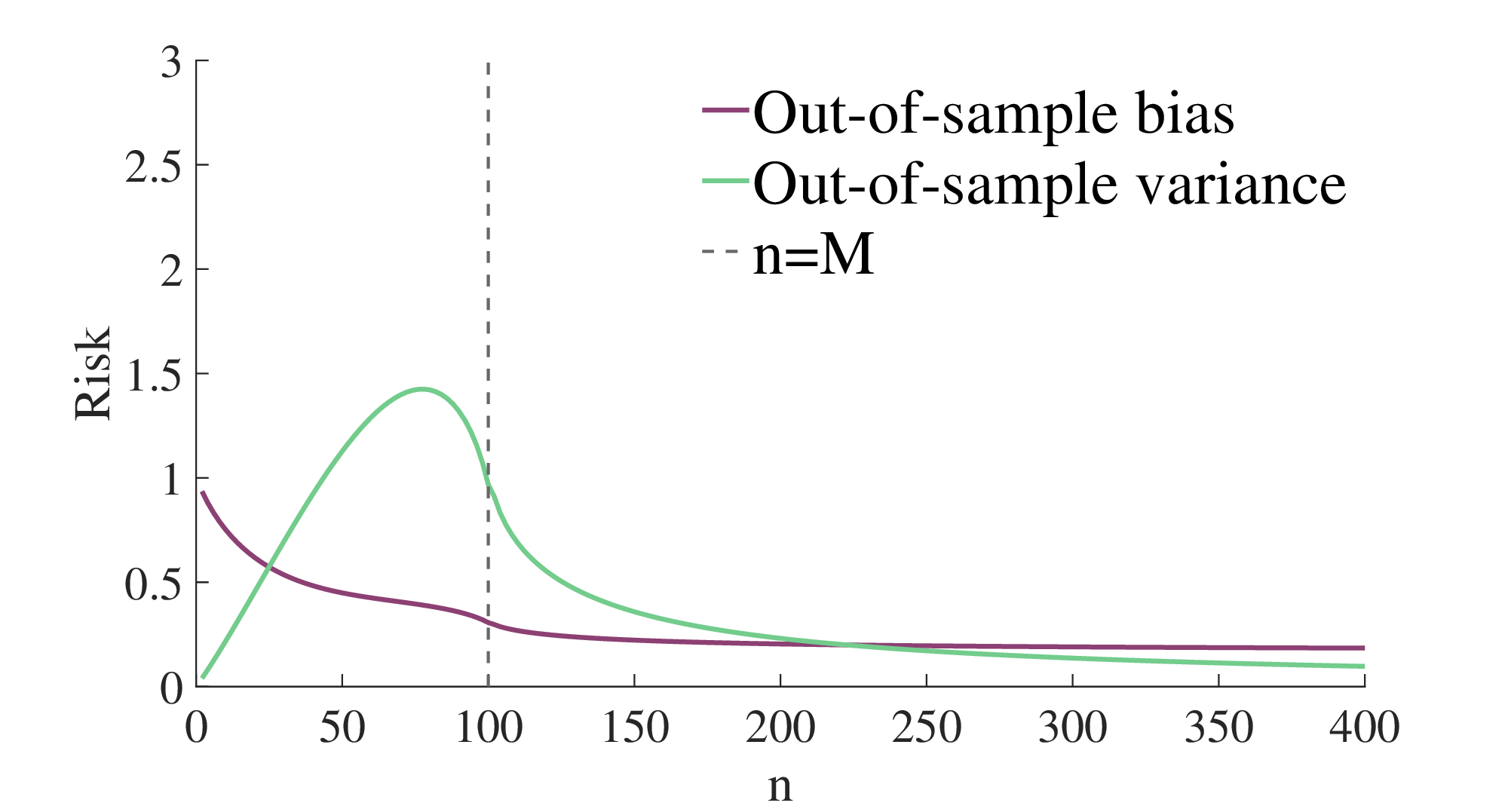}}
	\end{minipage}
	\vspace{-1em}
	\caption{Under equal weights, the limiting behaviors of out-of-sample bias $R_{\text{out},B}(\bm{\omega})$ and out-of-sample variance $R_{\text{out},V}(\bm{\omega})$ as $M$ and $n$ vary. The experimental setting is the same as in Figure~\ref{fig: risk_infty}.}
	\label{fig: bias var risk}
\end{figure*}

To analyze the global risk surface, we artificially exclude singular sub-models with $k_q = n$ from the candidate set when $M \ge n$.
Figure \ref{fig: risk} shows that the asymptotic risk surface of equally weighted model averaging exhibits a ``double descent'' phenomenon, with a reduced but still present risk peak compared to the single-model baseline.
The overall risk profile is driven by the interplay between out-of-sample bias $R_{\text{out},B}(\bm{\omega})$ and out-of-sample variance $R_{\text{out},V}(\bm{\omega})$, which exhibit disparate behaviors as shown in Figure \ref{fig: bias var risk}. 
We analyze this mechanism through two cross-sections:
\begin{figure*}[] 
	\centering 
	\begin{minipage}{0.64\linewidth}
		\centering
		\subfloat[]{\includegraphics[width=\linewidth]{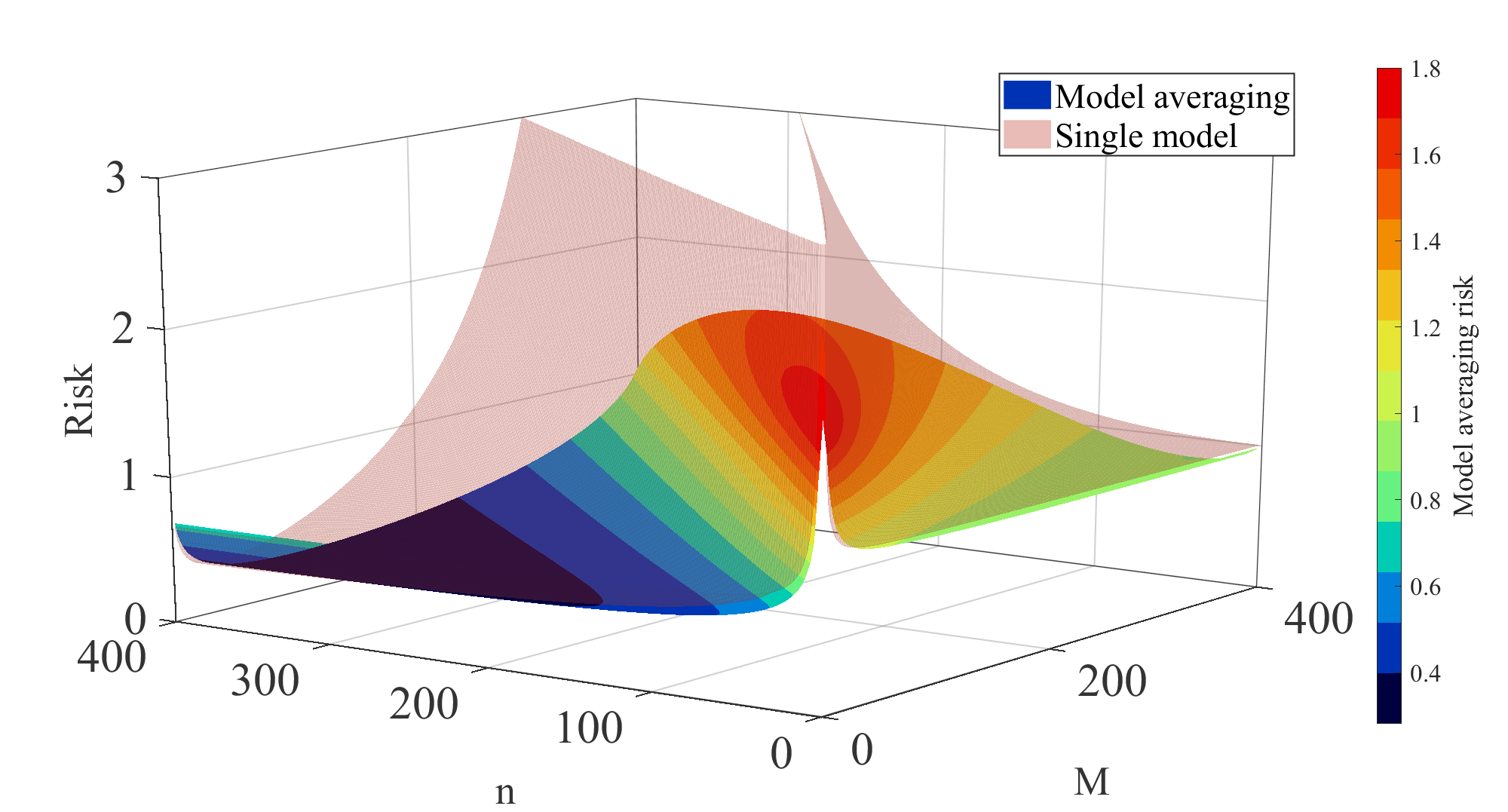}}
	\end{minipage}
	\begin{minipage}{0.35\linewidth}
		\centering
		\subfloat[fixed $n=100$]{\includegraphics[width=\linewidth]{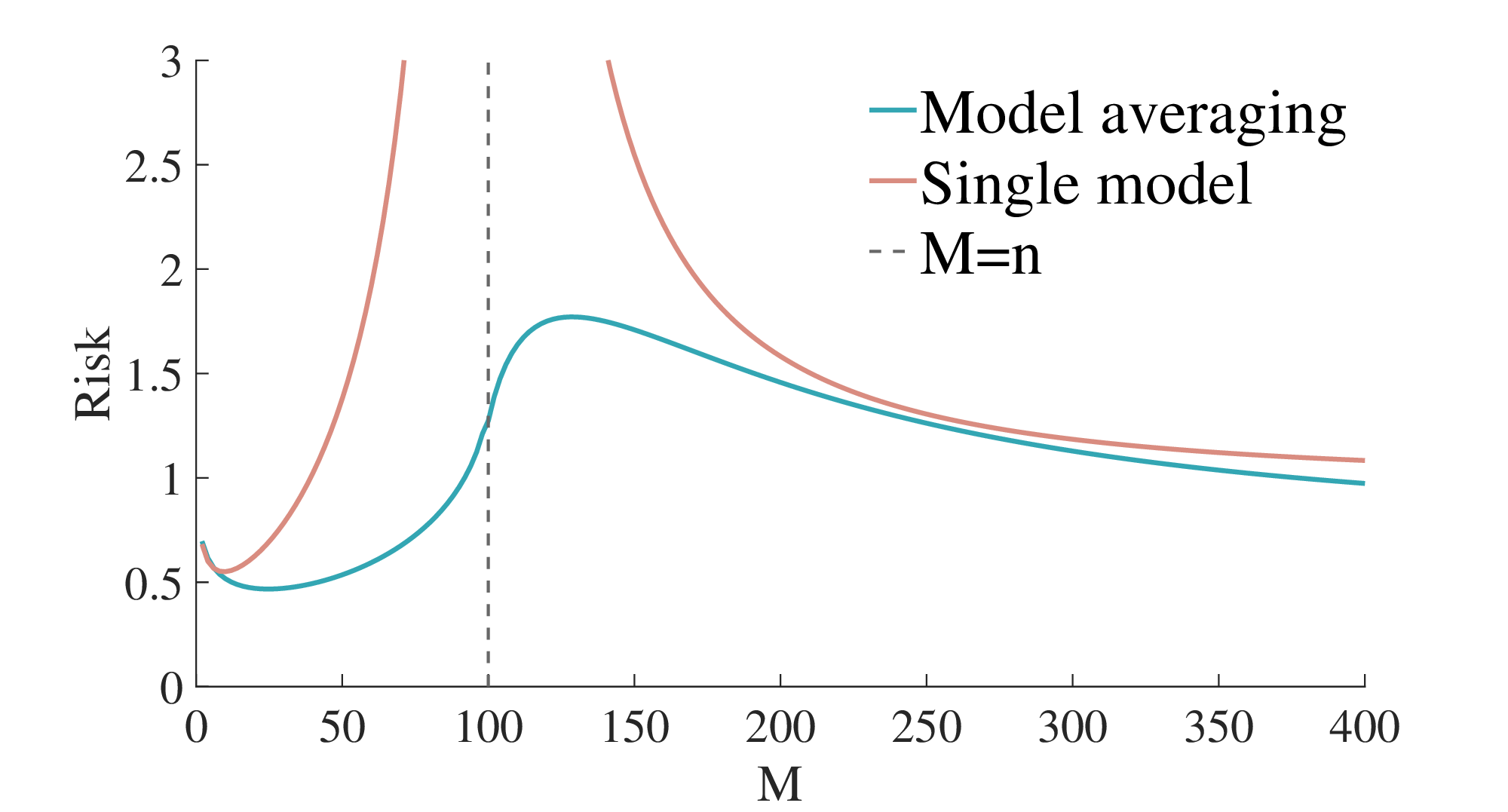}}\\
		\subfloat[fixed $M=100$]{\includegraphics[width=\linewidth]{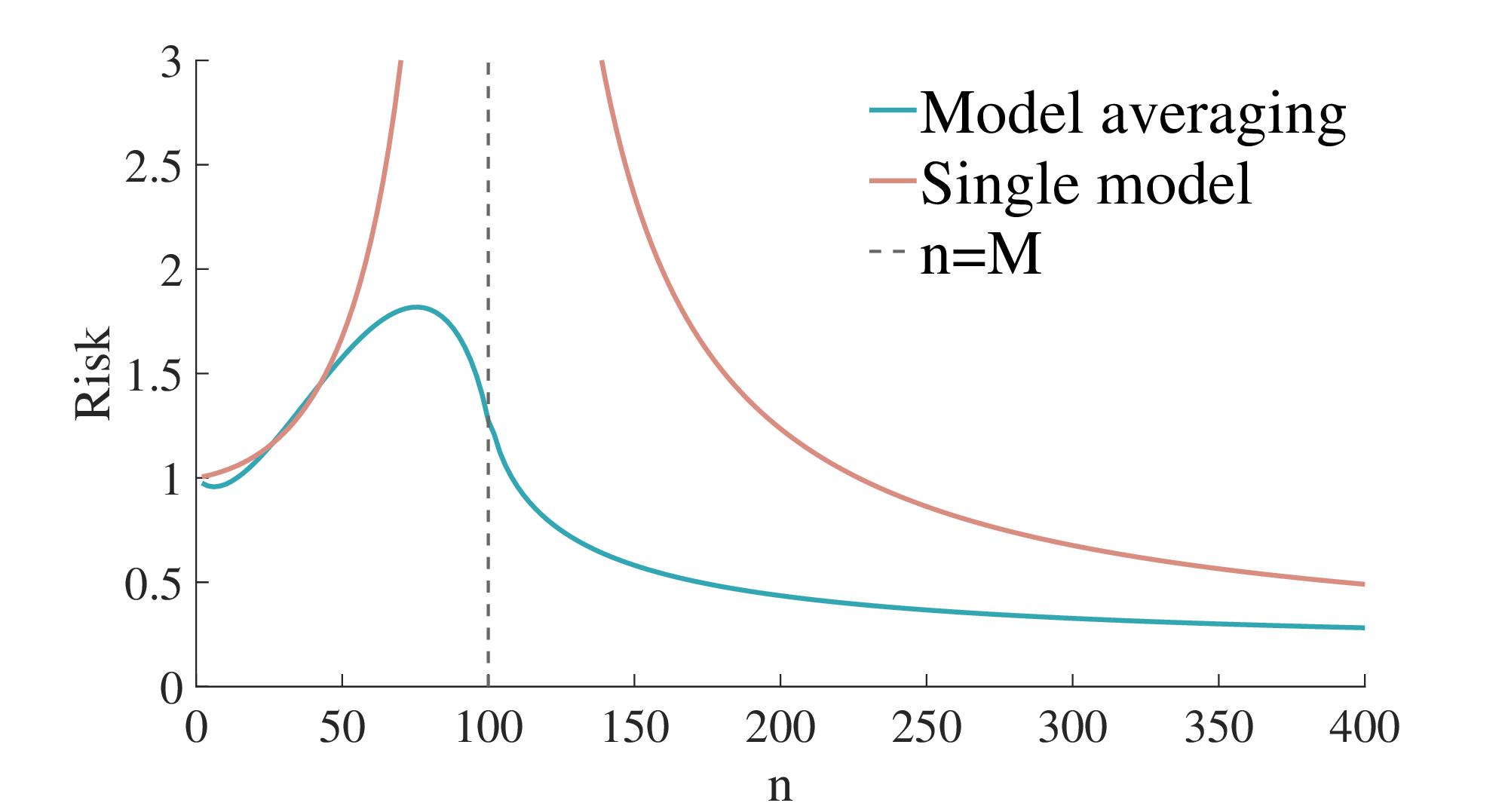}}
	\end{minipage}
	\vspace{-1em}
	\caption{Under equal weights (singular sub-models with $k_q=n$ excluded), the limiting behavior of out-of-sample risk $R_{\text{out}}(\bm{\omega})$ for model averaging and $R_{\text{single}}^{(q)}$ for a single model as $M$ and $n$ vary. The experimental setting is the same as in Figure~\ref{fig: risk_infty}.}
	\label{fig: risk}
\end{figure*}

Model-ensemble double descent (see Figures \ref{fig: risk}(b) and \ref{fig: bias var risk}(b)):
Fixing $n=100$, the risk traverses three regimes.
In the under-parameterized region ($M<n$),  
the classical bias-variance trade-off prevails; introducing models reduces bias while variance increases only moderately, driving an initial decline in ensemble risk.
However, near the interpolation boundary ($M = n$), incorporating models with dimensions approaching $n$ 
triggers a sharp variance cliff.
This extreme estimation variance overwhelms the bias reduction, forcing the overall risk to peak.
Subsequently, entering the over-parameterization ($M > n$), the implicit regularization of excess parameters takes effect.
The addition of high-dimensional sub-models mitigates the extreme variance via the equal-weight averaging mechanism, without sacrificing the previously achieved low bias, driving a secondary risk descent.

Sample-wise double descent (see Figures \ref{fig: risk}(c) and \ref{fig: bias var risk}(c)):
Fixing $M=100$ highlights the transition from data scarcity to abundance.	
In the data-starved regime ($n<M$), this is driven by the conflict between decreasing bias and inflated variance;
as $n$ approaches model dimensions, sub-models with estimation instability cause the variance to peak.
As $n$ surpasses $M$, the variance cliff vanishes (as all candidate models become under-parameterized). 
Both out-of-sample bias and variance then steadily decline, converging to a stable lower bound dictated by irreducible noise, consistent with classical large-sample theory.

\subsubsection{Ensemble emergence in strategically weighted model averaging}
While equal weighting can partially mitigate extreme variance, as hinted by \cite{NEURIPS2020_7d420e2b}, the ensemble risk peak near the interpolation boundary remains pronounced.
This observation is corroborated by our experiments with random weighting strategies (detailed in supplementary material), which similarly preserve the risk peak, resulting in a double-descent profile.
The underlying issue is that risk inflation is not confined solely to the singular model ($k_q = n$); the estimation variance of any sub-model whose dimension approaches the sample size diverges.
Because simple strategies assign indiscriminate weights to these unstable candidates, the ensemble inevitably inherits the extreme variance.

To suppress this risk peak, the weight allocation needs to root in principle: assigning smaller weights to high-risk models.
Leveraging our bias-variance decomposition, we note that the extreme risk peak around the interpolation boundary is driven by the out-of-sample variance; therefore, we naturally adopt the reciprocal of the asymptotic out-of-sample variance \eqref{var Dv} as a targeted risk penalty.
Specifically, the weight for the $q$th candidate model is
\begin{equation}\label{var_weight}
	\omega_q=\frac{1/\bm{D}_V(q,q)}{\sum_{m=1}^M 1/\bm{D}_V(m,m)}, \quad q=1,\cdots,M.
\end{equation}
By penalizing high variance, this strategy suppresses the localized risk surges, allowing us to reconstruct the complete nested sequence, retaining even the singular model ($k_q=n$).

\begin{figure*}[h] 
	\centering 
	\begin{minipage}{0.64\linewidth}
		\centering
		\subfloat[]{\includegraphics[width=\linewidth]{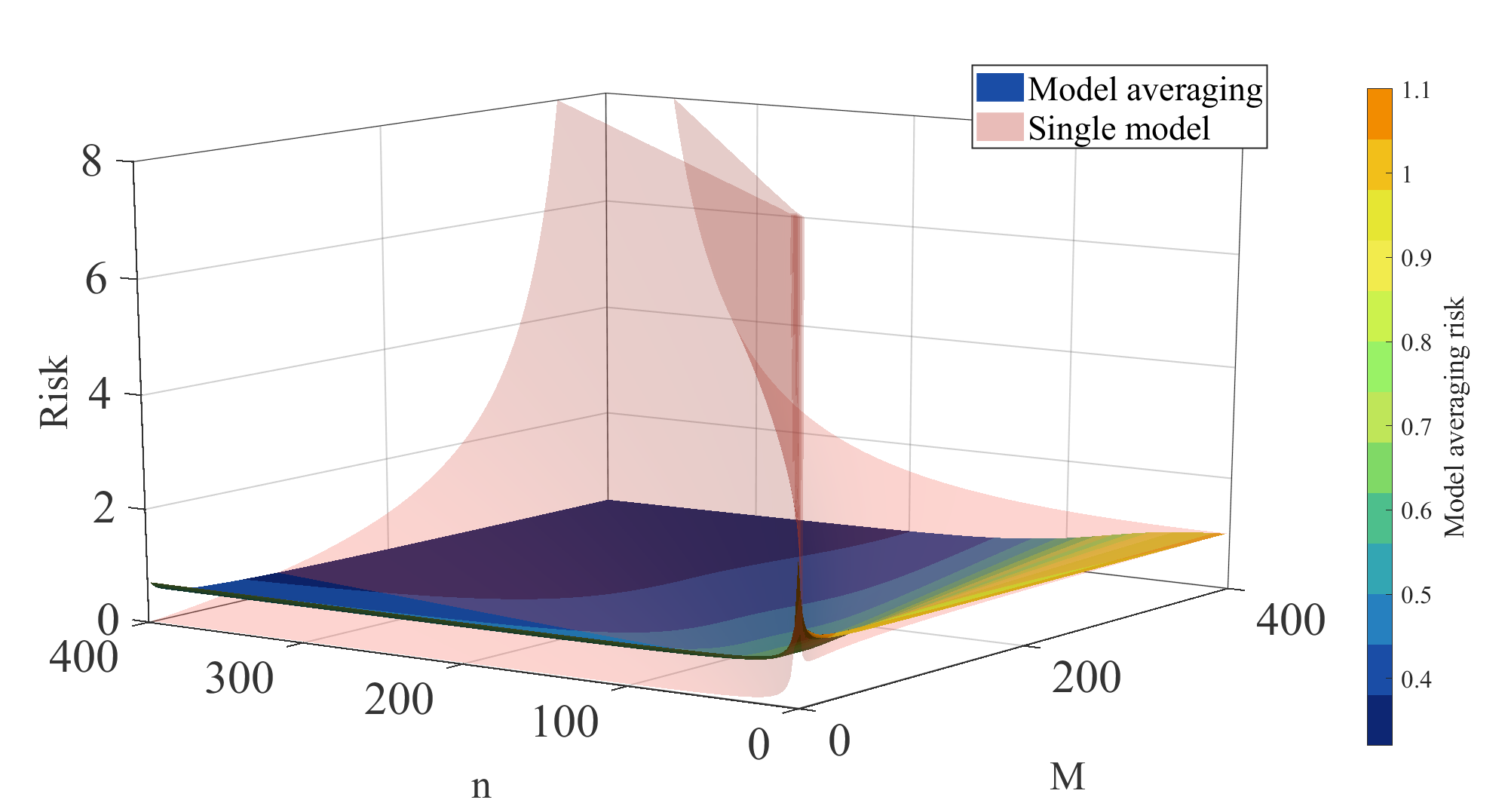}}
	\end{minipage}
	\begin{minipage}{0.35\linewidth}
		\centering
		\subfloat[fixed $n=100$]{\includegraphics[width=\linewidth]{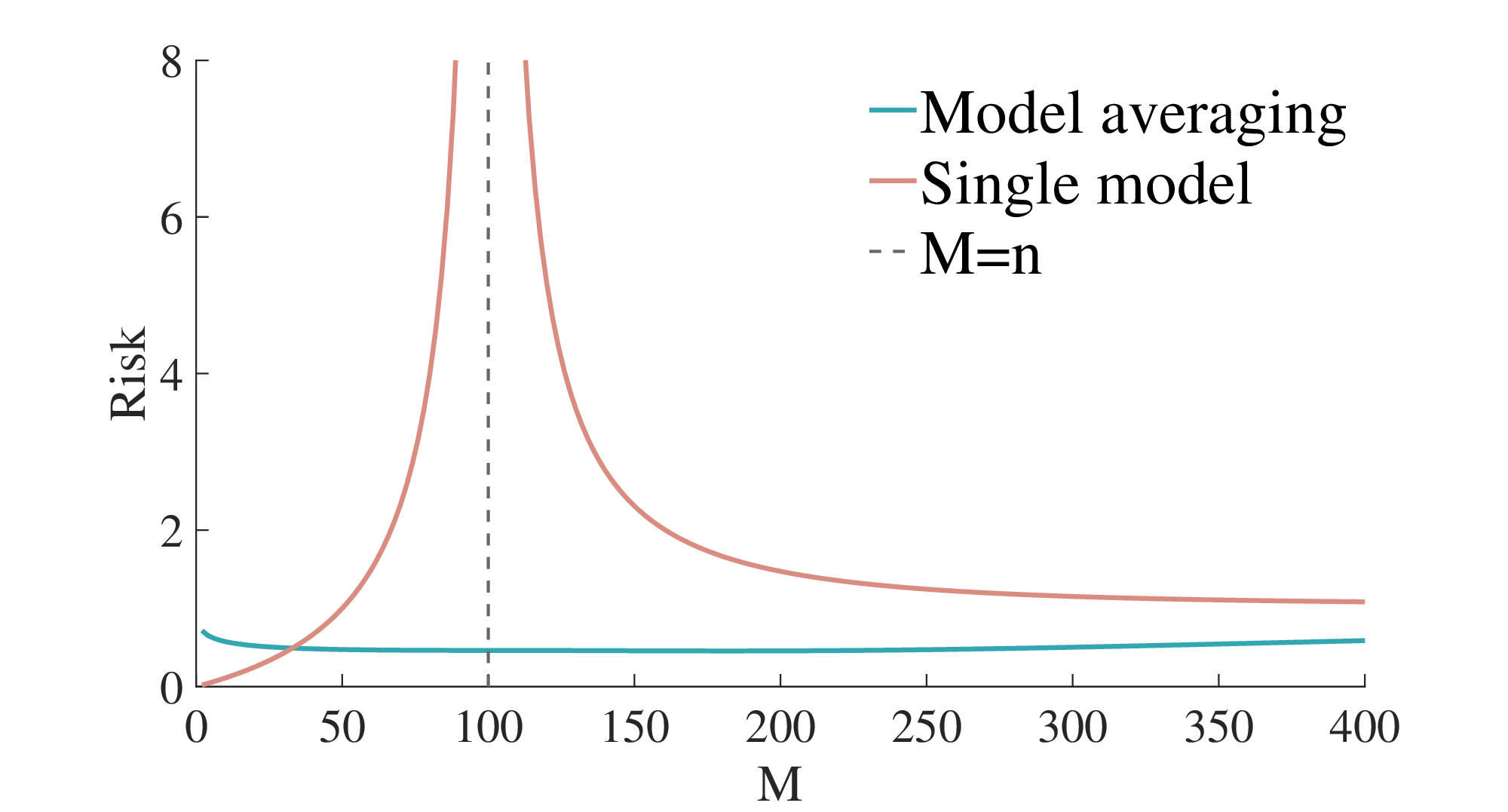}}\\
		\subfloat[fixed $M=100$]{\includegraphics[width=\linewidth]{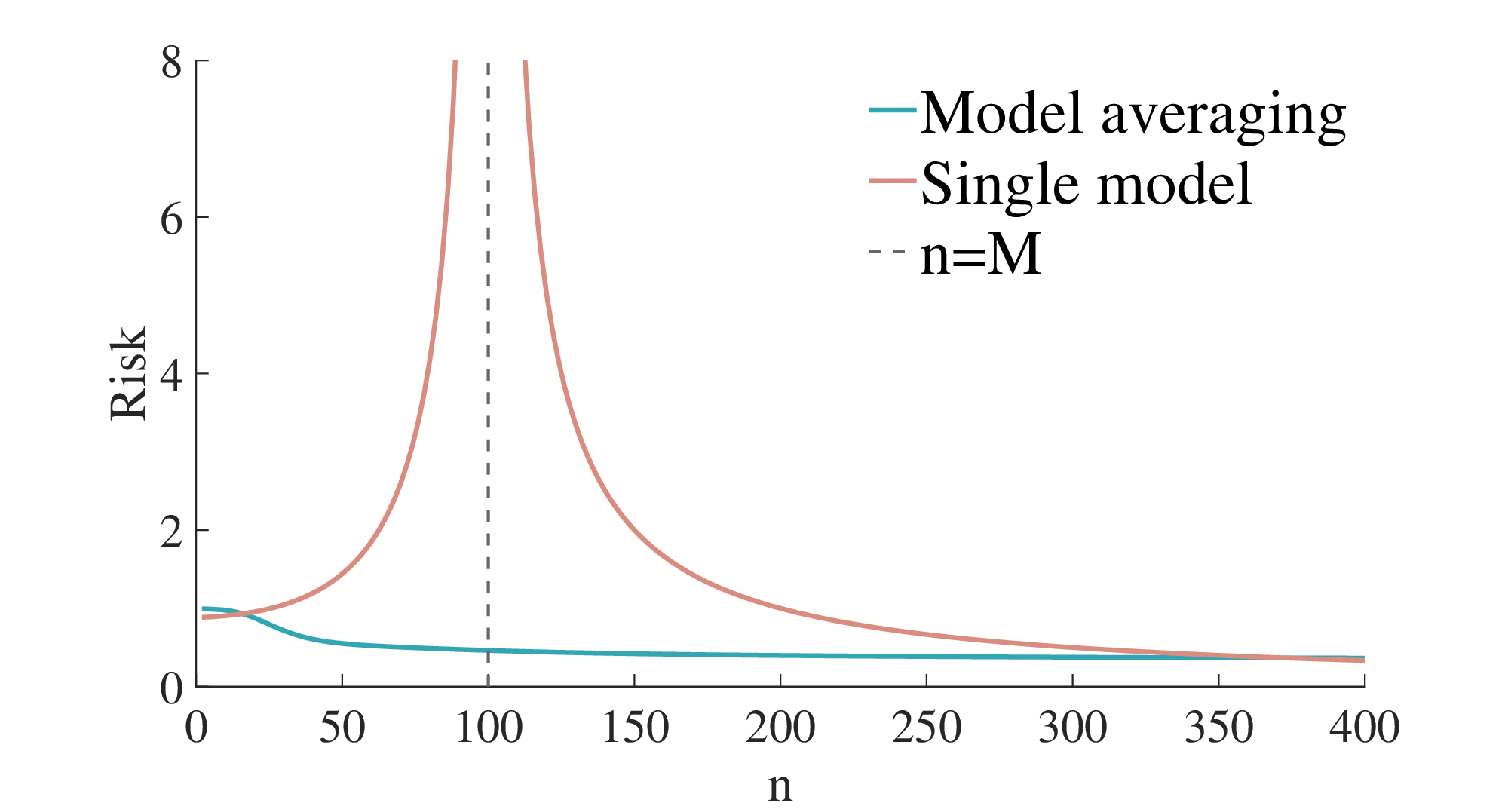}}
	\end{minipage}
	\vspace{-1em}
	\caption{Under variance-penalized weights \eqref{var_weight} (singular sub-models with $k_q=n$ retained), the limiting behaviors of out-of-sample risk $R_{\text{out}}(\bm{\omega})$ for model averaging and $R_{\text{single}}^{(q)}$ for a single model as $M$ and $n$ vary. The experimental setting is the same as in Figure~\ref{fig: risk_infty}.}
	\label{fig: emergence}
\end{figure*}

Figure \ref{fig: emergence} presents a three-dimensional risk surface comparison between our variance-penalized model averaging, $R_{\text{out}}(\bm{\omega})$, and
the risk of a single baseline model, $R_{\text{single}}$.  
The results reveal a clear contrast: 
as the dimension-to-sample ratio approaches the interpolation boundary ($M/n\to 1$), the risk surface of the single model exhibits a divergent double-descent ridge due to the singularity of the sample covariance matrix (\cite{10.1214/21-AOS2133}).
Conversely, the risk surface of the variance-penalized model averaging is globally flattened, maintaining a smooth trend across the entire $(n, M)$ space.

This pronounced flattening characterizes ensemble emergence, reminiscent of Anderson's principle that ``more is different'' (\cite{doi:10.1126/science.177.4047.393}).
Such a flat risk surface is absent in any isolated single model, which  suffers generalization collapse near the boundary $M=n$. 
When these candidates are combined via the variance-penalized mechanism, the ensemble diminishes the influence of near-singular models and diverts weight to well-conditioned candidates, thereby flattening the risk peak.
These findings suggest that the ``high-risk, low-weight'' principle facilitates this ensemble emergence by producing a globally flat risk.

\subsection{Extension to the case of general covariance matrices}\label{sec: general}
To accommodate real-world correlation structures, we relax the assumption of an identity covariance matrix to a general symmetric positive definite covariance matrix $\bm{\Sigma}$.
However, the exact out-of-sample risk depends heavily on $\bm{X}$ and $\bm{\Sigma}$, making direct weight optimization intractable.
The following theorem establishes the computable asymptotic expressions for both the bias and variance components.
\begin{theorem}\label{Th: trace}
	Consider a sequence of $M$ nested candidate models with dimensions $k_1 < k_2 < \dots < k_M$ and suppose that Assumption \ref{assump1} is satisfied.
	For any given weight vector $\bm{\omega} \in \mathcal{H}_n$, as $n, k_q \to \infty$ such that $k_q/n \to c_q \in (0, 1)$, we have
	$$
	\mathbb{E}\left[R_{\text{out},B}(\bm{\omega}) \right] \to \bm{\omega}^\prime\bm{B}_{\text{out}} \bm{\omega}
	\quad\text{and}\quad
	R_{\text{out},V}(\bm{\omega}) \xrightarrow{a.s.} \bm{\omega}^\prime \bm{V}_{\text{out}} \bm{\omega},
	$$
	where $\bm{B}_{\text{out}}$ and $\bm{V}_{\text{out}}$ are symmetric matrices whose $(q,l)$th entries are respectively given by
	\begin{equation*}
		\bm{B}_{\text{out}}(q,l) = \frac{1}{1-\min\{c_q, c_l\}} \phi_{\max\{q, l\}}
		\quad \text{and} \quad
		\bm{V}_{\text{out}}(q,l) = \sigma^2 \frac{\min\{c_q, c_l\}}{1-\min\{c_q, c_l\}},
	\end{equation*}
	with $\phi_q=\bm{\Theta}_{re(q)}^\prime\left(\bm{\Sigma}_{re(q),re(q)}-\bm{\Sigma}_{re(q),q} \bm{\Sigma}_{q}^{-1}\bm{\Sigma}_{q,re(q)}\right)\bm{\Theta}_{re(q)}, q=1,\cdots,M$. 
	Furthermore, if $\bm{X}\sim \mathcal{N}(\bm{0},\bm{\Sigma})$ and the signal strength satisfies $\bm{\Theta}^\prime\bm{\Sigma}\bm{\Theta}=o(\sqrt{n/ \log n })$, then the out-of-sample bias holds that
	\begin{equation*}
		R_{\text{out},B}(\bm{\omega}) \xrightarrow{a.s.} \bm{\omega}^\prime\bm{B}_{\text{out}} \bm{\omega} .
	\end{equation*}
\end{theorem}	
These limits transform sample-dependent matrix expressions into concise closed-form solutions depending solely on macroscopic parameters: 
the ratio $c_q$, the noise variance $\sigma^2$, and the signal strength $\phi_q$. 
This tractability allows us to directly embed the asymptotic out-of-sample risk into the weight optimization criterion, overcoming the fundamental limitations of traditional methods in high-dimensional scenarios.

\section{Large model averaging}
Building on these theoretical insights, we now construct a new weight choice criterion.
We begin by recalling the bias-variance decomposition of out-of-sample risk introduced in \eqref{decompose of R}.
For linear prediction $\hat{\mu}_0(\bm{\omega}) = \bm{x}_0^\prime \hat{\bm{\Theta}}(\bm{\omega})$
and true mean $\mu_0 = \bm{x}_0^\prime \bm{\Theta}$, the squared loss in \eqref{Riskout} is
$(\bm{x}_0^\prime \hat{\bm{\Theta}}(\bm{\omega}) - \bm{x}_0^\prime \bm{\Theta})^2$.
Because the new observation $\bm{x}_0$ is independent of the training data $\bm{X}$ with population covariance $\bm{\Sigma} = \mathbb{E}\left[\bm{x}_0 \bm{x}_0^\prime\right]$, taking the expectation with respect to $\bm{x}_0$ yields
$
\mathbb{E}\left[\left(\bm{x}_0^\prime (\hat{\bm{\Theta}}(\bm{\omega}) - \bm{\Theta})\right)^2 \;\middle|\; \bm{X},\bm{Y}\right]
= \left\|\hat{\bm{\Theta}}(\bm{\omega}) - \bm{\Theta}\right\|_{\bm{\Sigma}}^2 ,
$
where $\|\bm{a}\|_{\bm{\Sigma}}^2=\bm{a}^\prime\bm{\Sigma}\bm{a}$ denotes the squared norm induced by $\bm{\Sigma}$.
Hence, the out-of-sample risk \eqref{Riskout} can be expressed as
\begin{equation}\label{Rout_b_v}
	R_{\text{out}}(\bm{\omega})=\mathbb{E}\left[  L_{\text{out}}(\bm{\omega})  \mid \bm{X} \right]=\mathbb{E}\left[  \left\|\hat{\bm{\Theta}}(\bm{\omega}) - \bm{\Theta}\right\|_{\bm{\Sigma}}^2 \;\middle|\; \bm{X} \right]=R_{\text{out},B}(\bm{\omega})+R_{\text{out},V}(\bm{\omega}).
\end{equation}
Similarly, the in-sample risk decomposes as
\begin{equation}\label{Rin_b_v}
	R_{\text{in}}(\bm{\omega})=\mathbb{E}\left[  L_{\text{in}}(\bm{\omega})  \mid \bm{X} \right]=\mathbb{E}\left[  \left\|\hat{\bm{\Theta}}(\bm{\omega}) - \bm{\Theta}\right\|_{\widehat{\bm{\Sigma}}}^2 \;\middle|\; \bm{X} \right]=R_{\text{in},B}(\bm{\omega})+R_{\text{in},V}(\bm{\omega}),
\end{equation}
where $\widehat{\bm{\Sigma}} = \bm{X}^\prime\bm{X}/n$ is the sample covariance matrix, the in-sample bias is
\begin{equation}\label{RinBias}
	R_{\text{in},B}(\bm{\omega})\coloneqq \frac{1}{n}\sum_{i=1}^n \left(\mathbb{E}\left[\hat{\mu}_i(\bm{\omega}) \mid \bm{X}\right]-\mu_i\right)^2= \frac{1}{n}\|(\bm{I}-\bm{P}(\bm{\omega}))\bm{X}\bm{\Theta}\|^2,
\end{equation}
and the in-sample variance is 
\begin{equation}\label{RinVar}
	R_{\text{in},V}(\bm{\omega})\coloneqq\frac{1}{n}\sum_{i=1}^n\operatorname{Var}\left(\hat{\mu}_i(\bm{\omega}) \mid \bm{X}\right)=
	\frac{\sigma^2}{n}\operatorname{tr}\left(\bm{P}(\bm{\omega})^2\right).
\end{equation}
Here, $\hat{\bm{\mu}}(\bm{\omega})=\left(\hat{\mu}_1(\bm{\omega}),\cdots,\hat{\mu}_n(\bm{\omega})\right)^\prime$ denotes the vector of fitted values and $\bm{P}(\bm{\omega})$ is the corresponding weighted projection matrix.

The fundamental divergence between these two risks \eqref{Rout_b_v} and \eqref{Rin_b_v} lies in the quadratic matrix used to measure the loss: 
$R_{\text{in}}(\bm{\omega})$ relies on the sample covariance $\widehat{\bm{\Sigma}}$, while $R_{\text{out}}(\bm{\omega})$ depends on the population covariance $\bm{\Sigma}$. 
In the low-dimensional regime ($p/n \to 0$), $\widehat{\bm{\Sigma}}$ is a consistent estimator of $\bm{\Sigma}$, making the two risks asymptotically equivalent.  
However, in the high-dimensional regime ($p/n \to c \in(0, 1)$), the eigenvalues of $\widehat{\bm{\Sigma}}$ diverge significantly from those of $\bm{\Sigma}$.
Even in the isotropic case where $\bm{\Sigma} = \bm{I}_{p}$, the Mar\v{c}enko-Pastur law \cite{SILVERSTEIN1995331} indicates that the empirical spectral distribution of the sample covariance matrix $\widehat{\bm{\Sigma}}$ is supported on a wide interval $[(1-\sqrt{c})^2, (1+\sqrt{c})^2]$.
This spectral dispersion drives a divergence between $R_{\text{in}}(\bm{\omega})$ and $R_{\text{out}}(\bm{\omega})$:
The near-zero eigenvalues of $\widehat{\bm{\Sigma}}$ amplify noise during the fitting process, but when calculating in-sample loss, the estimation errors are weighted by those same eigenvalues, thereby masking the overfitting. 
The out-of-sample loss reweights these errors by $\bm{\Sigma}$, which does not contain these near-zero eigenvalues, thus exposing the overfitting and causing a sharp increase in the loss.

Consequently, traditional criteria targeting in-sample risk, such as MMA, become suboptimal for high-dimensional prediction. 
The MMA criterion is given by:
\begin{equation}\label{qth var hat}
	C_{\text{MMA}}(\bm{\omega})=	\frac{1}{n}\bm{\omega}^\prime  \bar{\bm{e}}^\prime\bar{\bm{e}}\bm{\omega}+\sum_{q=1}^M 2\omega_q \operatorname{tr}\left(\operatorname{Cov}\left(\hat{\bm{\Theta}}_{q}\;\middle|\;\bm{X}\right)\widehat{\bm{\Sigma}}_q\right),
\end{equation}
where $\bar{\bm{e}}=(\hat{\bm{e}}_{1},...,\hat{\bm{e}}_{M})$ is an $n\times M$ residual matrix. 
While $C_{\text{MMA}}(\bm{\omega})$ is unbiased for in-sample risk, i.e., $\mathbb{E}\left[C_{\text{MMA}}(\bm{\omega}) \mid \bm{X}\right]=R_{\text{in}}(\bm{\omega})+\sigma^2$, $R_{\text{in}}(\bm{\omega})$ ceases to be a valid surrogate for $R_{\text{out}}(\bm{\omega})$ when $k_q$ is comparable to $n$. 
For any given weight vector $\bm{\omega} \in \mathcal{H}_n$, the analytical expressions for the expected in-sample bias and in-sample variance,  referenced in \eqref{RinBias} and \eqref{RinVar}, are 
\begin{align}
	\mathbb{E}\left[R_{\text{in},B}(\bm{\omega}) \right] &=\sum_{q=1}^M \sum_{l=1}^M \omega_q \omega_l\frac{n-\max\{k_q, k_l\}}{n} \phi_{\max\{q, l\}}, \nonumber\\
	R_{\text{in},V}(\bm{\omega}) &=  \sigma^2\sum_{q=1}^M \sum_{l=1}^M \omega_q \omega_l\frac{\min\{k_q, k_l\}}{n}.\label{RinV=}
\end{align}
The calculation details are provided in \nameref{supp}.
Comparing these expressions with the asymptotic behavior of out-of-sample risk in Theorem~\ref{Th: trace} reveals that the in-sample risk systematically underestimates the true out-of-sample risk in expectation.  
Minimizing this misleading target, the criterion $C_{\text{MMA}}(\bm{\omega})$, leads to an excessive reliance on the training data, thereby compromising generalization performance.
Furthermore, the second term in \eqref{qth var hat} acts similarly to an $\ell_1$ regularization on the weights, scaled by the variance contribution of individual candidate models. 
This $\ell_1$-type penalty tends to  produce sparse weight solutions.
When candidate models carry complementary information, sparse weighting may discard useful signals, leading to increased bias or variance in the combined estimator. This not only undermines the robustness of the estimate but also directly degrades its generalization performance on new data.

To overcome these limitations and directly optimize out-of-sample generalization, we propose a novel weight choice framework, which is particularly effective when $k_M$ is comparable to $n$.
First, to address the systematic discrepancy between in-sample and out-of-sample variance in high-dimensional regimes, we introduce a variance correction term
\begin{equation}\label{Delta_V}
	\begin{aligned}
		\Delta_V(\bm{\omega})& \coloneqq R_{\text{out},V}(\bm{\omega})-R_{\text{in},V}(\bm{\omega})\\
		&\xrightarrow{a.s.} \bm{\omega}^\prime \bm{V}_{\text{out}}\bm{\omega}-\sigma^2\sum_{q=1}^M \sum_{l=1}^M \omega_q \omega_l \min\{c_q,c_l\}\\
		&=
		\sigma^2\sum_{q=1}^M \sum_{l=1}^M \omega_q \omega_l \frac{\min\{c_q,c_l\}^2}{1-\min\{c_q,c_l\}}>0
	\end{aligned}
\end{equation}
for $c_q,c_l\in (0,1)$.
The limiting behavior derived here is grounded in the asymptotic properties established in Theorem \ref{Th: trace} and \eqref{RinV=}.
This strictly positive limit reveals that the in-sample variance consistently underestimates the out-of-sample variance.
The magnitude of this underestimation exacerbates significantly when the ensemble includes candidate models whose number of regressors approaches the sample size (i.e., $c_q\to 1$).
By incorporating $\Delta_V(\bm{\omega})$ into the criterion $C_{\text{MMA}}(\bm{\omega})$, we mitigate the inherent over-optimism of in-sample metrics and ensure our weight choice is driven by actual out-of-sample performance.
In contrast, we do not apply a similar correction to the bias component. The bias discrepancy is defined as
\begin{small}
	\begin{align*}
		\Delta_B(\bm{\omega})\coloneqq R_{\text{out},B}(\bm{\omega})-R_{\text{in},B}(\bm{\omega})
		=\left(\mathbb{E}\left[\hat{\bm{\Theta}}(\bm{\omega})\;\middle|\; \bm{X}\right]-\bm{\Theta}\right)^\prime \left(\bm{\Sigma}-\widehat{\bm{\Sigma}}\right)\left(\mathbb{E}\left[\hat{\bm{\Theta}}(\bm{\omega})\;\middle|\; \bm{X}\right]-\bm{\Theta}\right).
	\end{align*}
\end{small}
Since $\Delta_B(\bm{\omega})$ depends directly on the unknown parameter $\bm{\Theta}$ and the degree of model misspecification, such correction for this bias component is practically infeasible.

Second, to mitigate potential information loss caused by sparse weighting and to enhance estimation robustness, we add a variance-weighted $\ell_2$ regularization term to $C_{\text{MMA}}(\bm{\omega})$: 
\begin{equation}\label{l2 regul}
	\xi\sum_{q=1}^M\omega_q^2 \operatorname{tr}\left(\operatorname{Cov}\left(\hat{\bm{\Theta}}_{q}\;\middle|\;\bm{X}\right)\bm{\Sigma}_q\right).
\end{equation}
This design serves two purposes: (i) it restricts extreme weights fluctuations, encouraging a smoother weight distribution that enables the estimator to incorporate information from all candidate models; (ii) by embedding the out-of-sample variance into the penalty weights, it imposes stronger shrinkage on high-variance models, thereby reducing the overall out-of-sample variance and enhancing generalization performance.
Combining these components \eqref{Delta_V} and  \eqref{l2 regul} yields the complete theoretical criterion of LaMA:
\begin{equation}\label{our MA}
	C_{\text{LaMA}}(\bm{\omega})= 	\underbrace{C_{\text{MMA}}(\bm{\omega})}_{\text{ estimator of in-sample risk}}\ +\ \underbrace{\Delta_V(\bm{\omega})}_{\text{variance correction}} \ +\ \underbrace{\xi\ \sum_{q=1}^M  \omega_q^2 \operatorname{tr}\left(\operatorname{Cov}\left(\hat{\bm{\Theta}}_{q}\;\middle|\;\bm{X}\right)\bm{\Sigma}_q\right)}_{\text{adaptive $\ell_2$ penalty} }.
\end{equation}

However, criterion \eqref{our MA} involves unknown population quantities and cannot be directly computed. To obtain a feasible criterion, we employ Theorem \ref{Th: trace} to replace these unknown terms with their asymptotic limits. Specifically, following \eqref{Delta_V}, we estimate $\Delta_V(\bm{\omega})$ by 
\begin{align*}
	\hat{\Delta}_V(\bm{\omega})&=\bm{\omega}^\prime \hat{\bm{V}}_{\text{out}} \bm{\omega}-\bm{\omega}^\prime \hat{\bm{V}}_{\text{in}}\bm{\omega},
\end{align*}
where $\hat{\bm{V}}_{\text{out}}$ and $\hat{\bm{V}}_{\text{in}}$ are estimators of $\bm{V}_{\text{out}}$ and $\bm{V}_{\text{in}}$, with the theoretical limit $c_q$ and the noise variance $\sigma^2$ replaced by their empirical counterpart $k_q/n$ and estimator $\hat{\sigma}^2$, respectively. 
In practice, the noise variance can be estimated by $\hat{\sigma}^2=\hat{\bm{e}}_K^\prime\hat{\bm{e}}_K/(n-k_K)$, where $K$ indexes a ``large'' candidate model with $k_K$ parameters (see \cite{MMA2007}); other estimation methods can also be chosen. 
By substituting \eqref{qth var hat} and the aforementioned estimators of parameters into the theoretical criterion \eqref{our MA}, we construct the empirical LaMA criterion
\begin{equation}\label{optimi}
	\hat{C}_{\text{LaMA}}(\bm{\omega})= \frac{1}{n} \bm{\omega}^\prime  \bar{\bm{e}}^\prime\bar{\bm{e}}\bm{\omega}+2\hat{\sigma}^2\sum_{q=1}^M \omega_q \frac{k_q}{n} + \ \hat{\Delta}_V(\bm{\omega})\ + \xi \bm{\omega}^\prime  \operatorname{diag}(\hat{\bm{V}}_{\text{out}})\bm{\omega},
\end{equation} 
where $\operatorname{diag}(\hat{\bm{V}}_{\text{out}})$ represents the diagonal elements of matrix $\hat{\bm{V}}_{\text{out}}$.
Simple algebra shows that the criterion \eqref{optimi} can be equivalently reformulated as
\begin{equation}\label{LaMA=B+V}
	\begin{aligned}
		\hat{C}_{\text{LaMA}}(\bm{\omega})=&\frac{1}{n}\bm{\omega}^\prime  \bar{\bm{e}}^\prime\bar{\bm{e}}\bm{\omega}+2\hat{\sigma}^2 \sum_{q=1}^M \omega_q \frac{k_q}{n} +  \bm{\omega}^\prime \left(\hat{\bm{V}}_{\text{out}}-\hat{\bm{V}}_{\text{in}}\right) \bm{\omega} + \xi \bm{\omega}^\prime  \operatorname{diag}(\hat{\bm{V}}_{\text{out}})\bm{\omega} \\
		=&\bm{\omega}^\prime \hat{\bm{B}}_{\text{in}} \bm{\omega} + \bm{\omega}^\prime \hat{\bm{V}}_{\text{out}} \bm{\omega} + \xi \bm{\omega}^\prime  \operatorname{diag}(\hat{\bm{V}}_{\text{out}})\bm{\omega}
	\end{aligned}
\end{equation}
for any given weight vector $\bm{\omega} \in \mathcal{H}_n$, where $\hat{\bm{B}}_{\text{in}}$ is an $M \times M$ matrix with its quadratic form $$\bm{\omega}^\prime \hat{\bm{B}}_{\text{in}} \bm{\omega}= \frac{1}{n}\bm{\omega}^\prime  \bar{\bm{e}}^\prime\bar{\bm{e}}\bm{\omega}+\hat{\sigma}^2 \sum_{q=1}^M\sum_{l=1}^{M} \omega_q \omega_l \frac{\max\{k_q,k_l\}}{n}.$$
To see the connection with in-sample bias, note that $R_{\text{in},B}(\bm{\omega})$ in \eqref{RinBias} can be expressed as 
\begin{align*}
	R_{\text{in},B}(\bm{\omega})&=\mathbb{E}\left[\frac{1}{n}\left\|\bm{Y}-\hat{\bm{\mu}}(\bm{\omega})\right\|^2
	\;\middle|\; \bm{X}\right]-\mathbb{E}\left[\frac{1}{n}\left\| (\bm{I}-\bm{P}(\bm{\omega}))\bm{e}\right\|^2 \;\middle|\;\bm{X}\right]\\
	&=\mathbb{E}\left[\frac{1}{n}\bm{\omega}^\prime  \bar{\bm{e}}^\prime\bar{\bm{e}}\bm{\omega}
	\;\middle|\; \bm{X}\right]-\sigma^2+\sigma^2\left(2\sum_{q=1}^M \omega_q \frac{k_q}{n}-\sum_{q=1}^M\sum_{l=1}^{M} \omega_q \omega_l \frac{\min\{k_q,k_l\}}{n}\right).
\end{align*}
Under the assumption that $\hat{\sigma}^2$ is an unbiased estimator of $\sigma^2$, the term $\bm{\omega}^\prime \hat{\bm{B}}_{\text{in}} \bm{\omega}-\hat{\sigma}^2$ serves as an unbiased estimator of in-sample bias.
Consequently, this decomposition reveals that the empirical LaMA criterion \eqref{LaMA=B+V} consists of three terms: an in-sample bias estimate, an out-of-sample variance estimate, and a regularization term.	
In this third term, the diagonal matrix $\operatorname{diag}(\hat{\bm{V}}_{\text{out}})$ captures the relative variance structure across candidate models, and the scalar $\xi$ calibrates the global regularization strength.

We propose a heuristic 
choice based on the principle that models with higher out-of-sample variance should incur stronger penalization.
Specifically, we set the regularization parameter $\xi$ to the maximum ratio of out-of-sample to in-sample variance:
\begin{equation}\label{regularization}
	\xi  = \max_{q\in \{1,\cdots,M\}}\frac{\hat{\bm{V}}_{\mathrm{out}}(q,q)}{\hat{\bm{V}}_{\mathrm{in}}(q,q)}.
\end{equation}
Similar to the regularization scaling strategy used in \cite{10.1214/17-AOS1549}, our chosen ratio captures the degree of variance inflation: a larger ratio indicates a larger discrepancy between in-sample and out-of-sample variance, which is a characteristic of high-variance models.
To mitigate the model averaging risk, we use $\xi$ as the regularization strength, so that penalization is scaled to match the degree of variance inflation.  
In particular, $\xi$ becomes large when the candidate set contains models approaching interpolation, thereby heavily penalizing and suppressing the weights assigned to high-variance models. 
Conversely, when all candidate models are far from the interpolation boundary, this ratio approaches $1$, and the penalty weakens accordingly, allowing the bias-variance trade-off (i.e., the first two terms in \eqref{LaMA=B+V}) to primarily determine the weight choice.
Finally, by substituting the chosen $\xi$ into \eqref{LaMA=B+V} and scaling the entire expression by the sample size $n$ to make the optimization problem more suitable for numerical calculation,
we develop a feasible LaMA method, which is summarized in Algorithm \ref{alg: RMT MA}. The weight $\omega$ of LaMA is obtained by
\begin{small} 
	\begin{equation}\label{MA optimi}
		\begin{aligned}
			\hat{\bm{\omega}} &= \mathop{\arg\min}_{\bm{\omega} \in \mathcal{H}_{n}} \
			n\,\hat{C}_{\text{LaMA}}(\bm{\omega})\\
			&= \mathop{\arg\min}_{\bm{\omega} \in \mathcal{H}_{n}} \ \left( \bm{\omega}^\prime  \bar{\bm{e}}^\prime\bar{\bm{e}}\bm{\omega}+\hat{\sigma}^2 \sum_{q=1}^M\sum_{l=1}^{M} \omega_q \omega_l \left(\max\{k_q,k_l\} + \frac{n \min\{k_q, k_l\}}{n-\min\{k_q, k_l\}}\right) + \xi \hat{\sigma}^2 \sum_{q=1}^M \omega_q^2\frac{nk_q}{n-k_q} \right).
		\end{aligned}
	\end{equation} 
\end{small}
The criterion innovatively combines the in-sample bias estimation and the asymptotic representations of out-of-sample variance, thereby simultaneously optimizing the fitting accuracy for known data and the generalization ability for unknown data.   
It breaks through the strict limitation of traditional methods that the model dimension is much smaller than the sample size ($k_M\ll n$).   
Its theoretical framework and algorithm design are applicable to scenarios where the ratio of the model dimension to the sample size tends to be a constant, providing a computationally efficient and easy-to-implement solution for practical applications.

\begin{algorithm}[htb]
	\caption{Large model averaging (LaMA).}\label{alg: RMT MA}
	\begin{algorithmic}[1]
		\Require The samples $\{\bm{X}_{n\times p},\bm{Y}_{n \times 1}\}$ (with the variables already ordered) and the number of regressors $\{k_1,k_2,\cdots,k_M\}$ for the candidate models.
		\Ensure The model averaging estimator of $\bm{\Theta}$.
		\State Calculate the estimator $\hat{\bm{\Theta}}_{q}$ by \eqref{MME} and the residual vector \(\hat{\bm{e}}_{q} = \bm{Y} - \bm{X}_{(q)}\hat{\bm{\Theta}}_{q}\);
		\State Construct the residual matrix $\bar{\bm{e}}=(\hat{\bm{e}}_{1},...,\hat{\bm{e}}_{M})$;
		\State Calculate the regularization parameter $\xi$ by \eqref{regularization};	
		\State Solve the optimization problem  \eqref{MA optimi};
		\State Model averaging estimator is $\hat{\bm{\Theta}}(\bm{\omega})=\left(
		\begin{array}{c c c}
			\hat{\bm{\Theta}}_{1} &\cdots&  \hat{\bm{\Theta}}_{M}\\
			\bm{0}_{(p-k_1)\times 1}  &\cdots & \bm{0}_{(p-k_M)\times 1}
		\end{array}\right)\hat{\bm{\omega}}$.
	\end{algorithmic}	
\end{algorithm}

\section{Simulation}

\subsection{Experimental setup and metrics}
Following the experimental setup of \cite{MMA2007}, the data are generated according to the linear model
\begin{equation}\label{generate model}
	y_i = \sum_{j=1}^{p} \theta_j x_{ij} + e_i, \quad i = 1,\dots,n,
\end{equation}
where the intercept term $x_{i1}=1$, the remaining regressors $x_{ij}, j=2,3,\dots$ are drawn from $\mathcal{N}(0,\bm{\Sigma})$, and the independent errors $e_i$ are from $\mathcal{N}(0,1)$. The regression coefficients are given by $\theta_j = c\sqrt{2\alpha}\, j^{-\alpha-1/2}$, where $c$ is determined by $R^2 = c^2/(1+c^2) \in \{0.1,\dots,0.9\}$, and the attenuation rate $\alpha=0.5$ controls the rate at which the coefficients decay. The experiment considers sample sizes $n \in \{25, 50, 150, 300\}$ and a fixed total dimension $p = 1000$. We construct three sets of nested candidate models, with the number of models $M \in \{\lfloor 3n^{1/3} \rceil, \lfloor 0.5n \rceil, \lfloor 0.9n \rceil\}$, where $\lfloor \cdot \rceil$ denotes rounding to the nearest integer and
the $q$th model contains the first $q$ regressors. Thus covering a wide range of scenarios with dimensions from much less than $n$ to close to $n$.

We compare the proposed method with several existing approaches: AIC model selection (AIC), smoothed AIC averaging (S-AIC), BIC model selection (BIC), smoothed BIC averaging (S-BIC), MMA, and jackknife model averaging (JMA).
To evaluate the predictive performance of the proposed LaMA method, we employ two metrics: relative in-sample loss and relative out-of-sample loss, defined as
$$\frac{\|\hat{\bm{\mu}}_{\text{train}}(\bm{\omega}) - \bm{\mu}_{\text{train}} \|^2}{\min_{1 \le q \le M} \|\hat{\bm{\mu}}_{q,\text{train}} - \bm{\mu}_{\text{train}} \|^2} \quad \text{and} \quad \frac{\|\hat{\bm{\mu}}_{\text{test}}(\bm{\omega}) - \bm{\mu}_{\text{test}} \|^2}{\min_{1 \le q \le M} \|\hat{\bm{\mu}}_{q,\text{test}} - \bm{\mu}_{\text{test}} \|^2},$$
respectively.	Here, $\{\bm{Y}_{\text{train}},\bm{X}_{\text{train}}\}$ and $\{\bm{Y}_{\text{test}},\bm{X}_{\text{test}}\}$ denote the independently generated training and test sets, where $\bm{\mu}$ and $\hat{\bm{\mu}}$ represent the true conditional means and their corresponding estimates. The index $q$ ranges over the $M$ candidate models.

\subsection{Performance analysis}
We conduct a series of simulation experiments under different dimensional regimes (the ratio of $M/n$) and different signal strengths ($R^2$). The aim is to study the performance of different model averaging methods during the transformation from low-dimensional to high-dimensional settings. The primary focus is on prediction accuracy and the sensitivity to noise.
In all comparisons, the error variance $\sigma^2$ is estimated from the model selected by BIC. Results obtained using the alternative variance estimator based on the largest candidate model are reported in supplementary material.
\begin{figure*}[htb]
	\centering
	\subfloat{\includegraphics[scale=0.35]{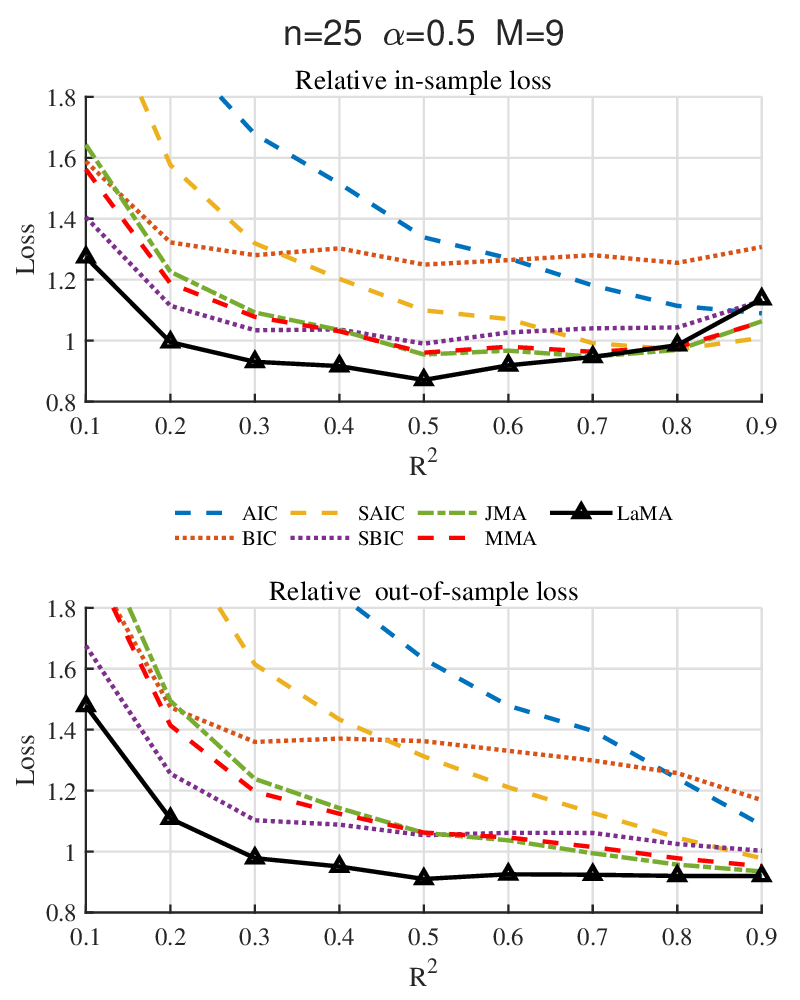}}	
	\subfloat{\includegraphics[scale=0.35]{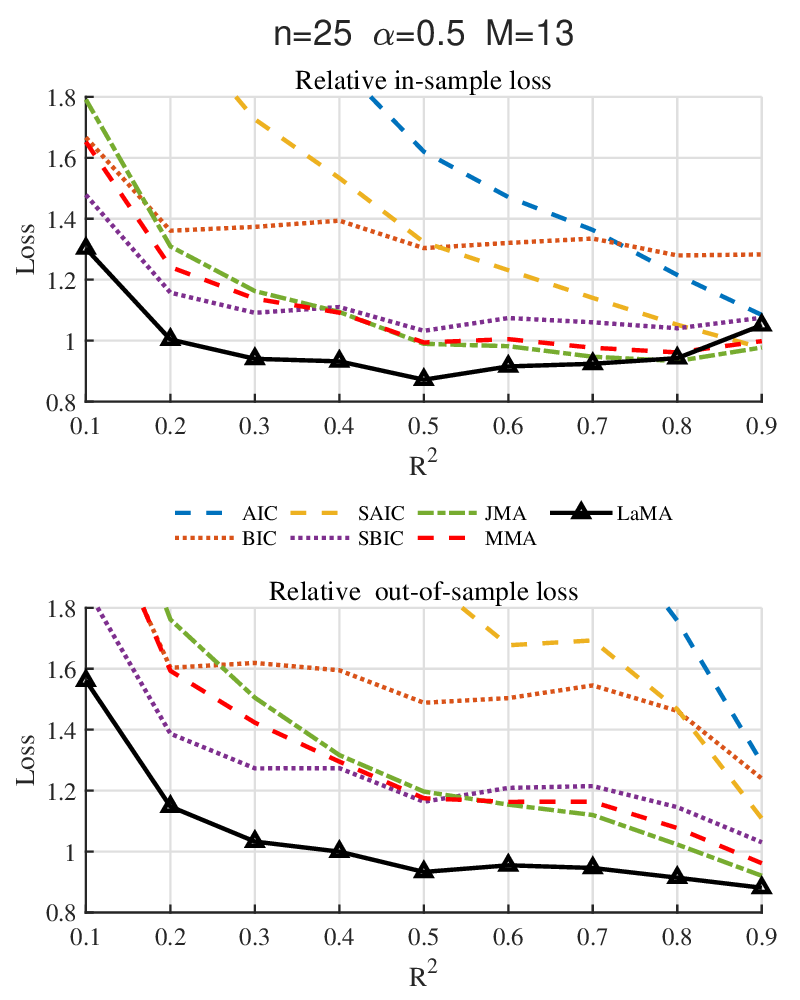}}
	\subfloat{\includegraphics[scale=0.35]{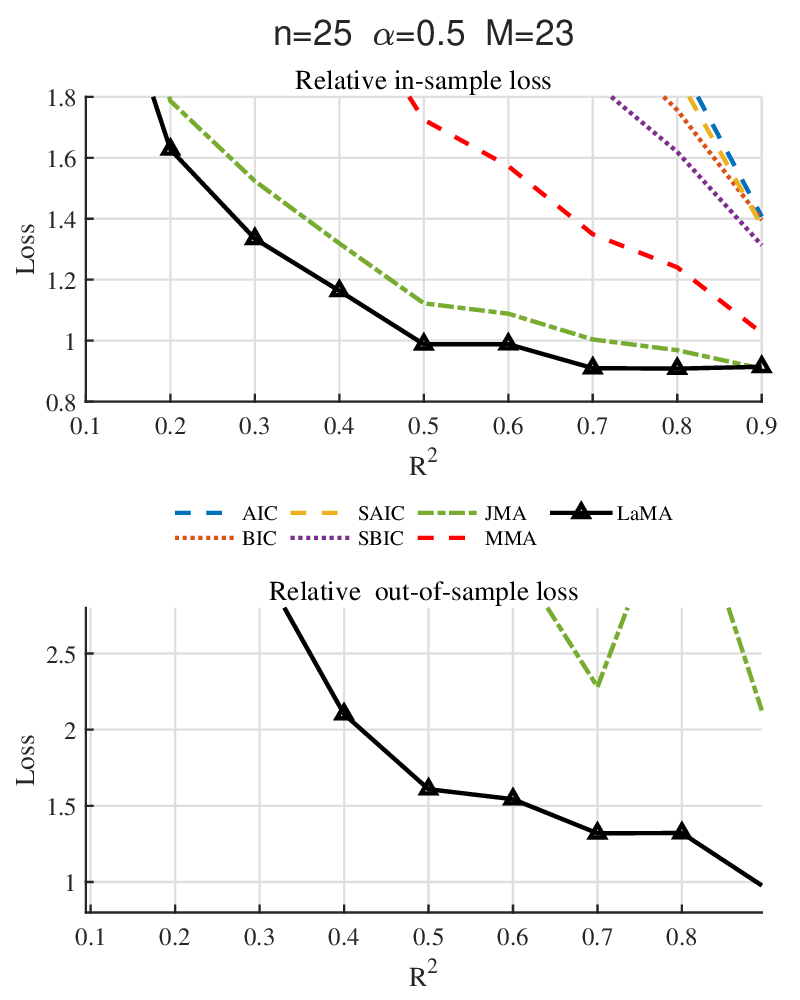}}
	\setlength{\abovecaptionskip}{0pt}
	\caption{ $n=25, \alpha=0.5, M\in \{ 9, 13, 23\}$.}
	\label{fig: n25}
\end{figure*}
\begin{figure*}[htb]
	\centering
	\subfloat{\includegraphics[scale=0.35]{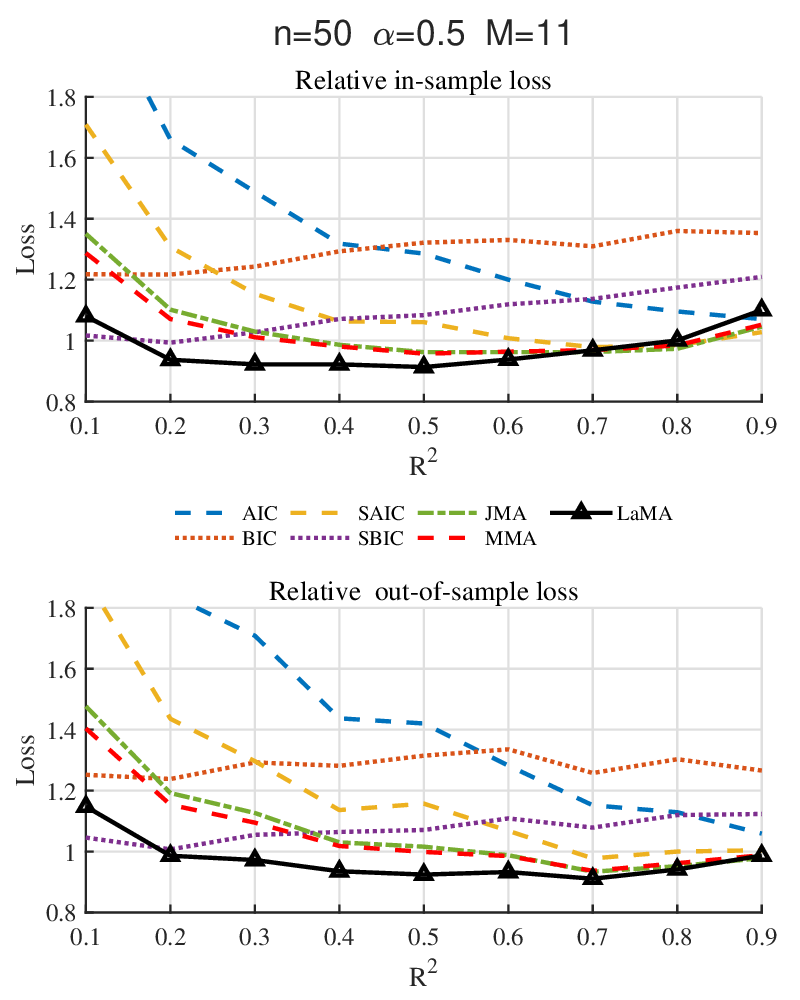}} \subfloat{\includegraphics[scale=0.35]{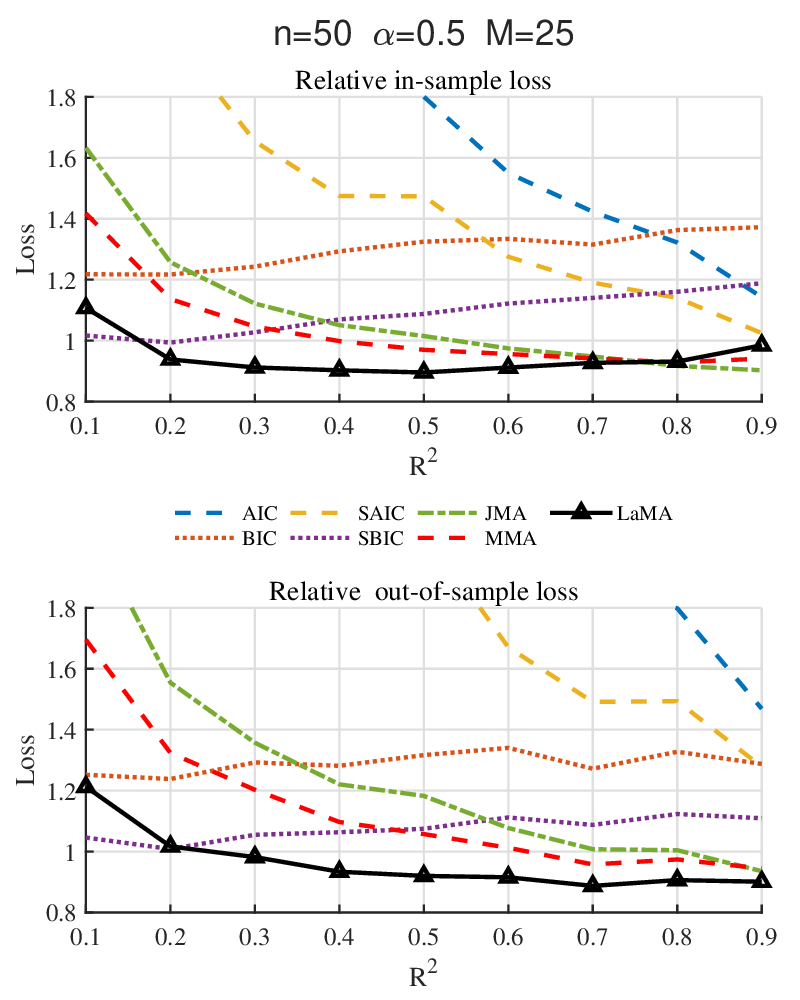}}
	\subfloat{\includegraphics[scale=0.35]{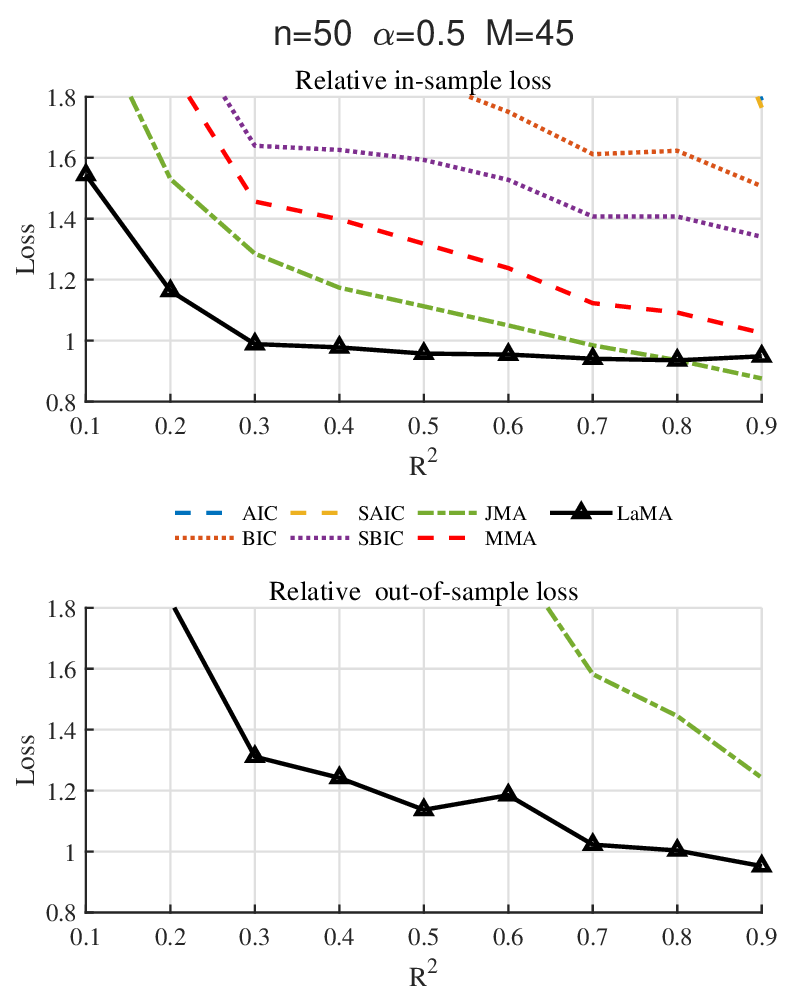}}
	\setlength{\abovecaptionskip}{0pt}
	\caption{ $n=50, \alpha=0.5, M\in \{ 11, 25, 45\}.$}
	\label{fig: n50}
\end{figure*}

Figure \ref{fig: n25} and Figure \ref{fig: n50} illustrate the results for relatively small sample sizes $(n = 25, 50)$. 
Under configurations with low-to-moderate model dimensions, LaMA achieves the lowest out-of-sample loss among all evaluated methods, with its performance advantage becoming more pronounced as model dimension increases.
However, a performance breakdown for all methods is observed when $n=25, M=23$ and $n=50, M=45$ under weak-signal (low $R^2$) settings.
This is because, in high-dimensional and small-sample scenarios, the asymptotic properties of traditional model selection and averaging methods fail to hold, leading to suboptimal predictive performance.
While competing methods exhibit diverging out-of-sample losses under these extreme settings, LaMA maintains the lowest loss curve. 
This relative advantage occurs because LaMA imposes a penalty on models with high out-of-sample variance, thereby achieving a more stable trade-off between bias and variance. This effectively mitigates the excessive fluctuations caused by complex models under small samples.
\begin{figure*}[htb]
	\centering
	\subfloat{\includegraphics[scale=0.35]{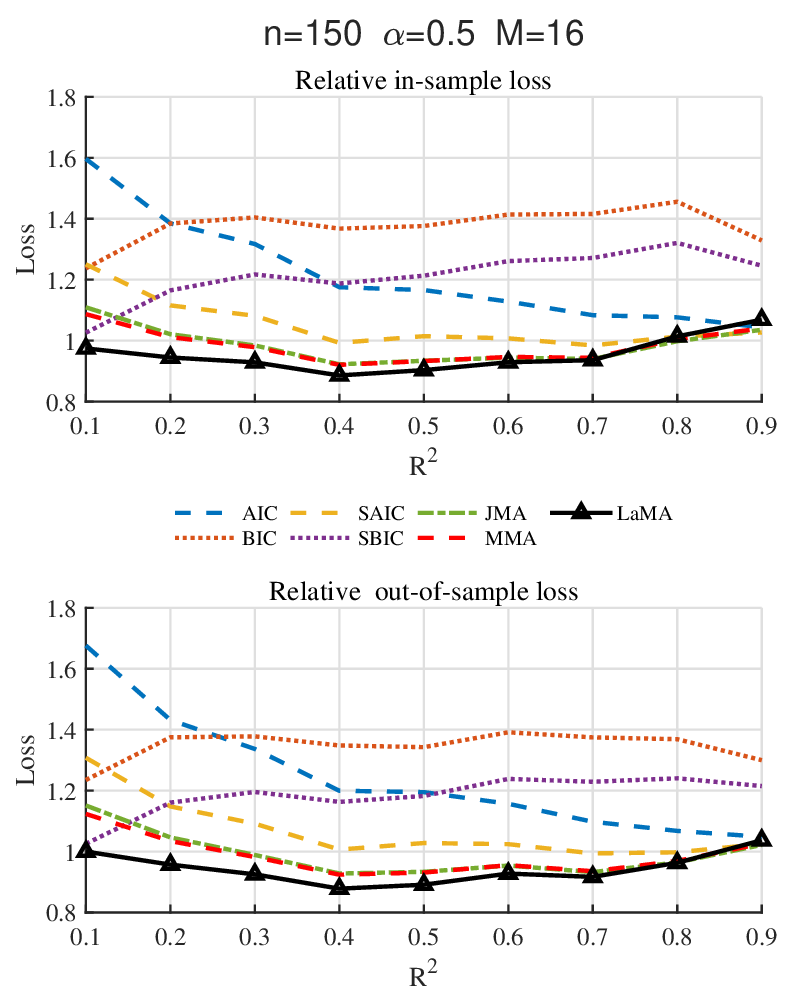}}
	\subfloat{\includegraphics[scale=0.35]{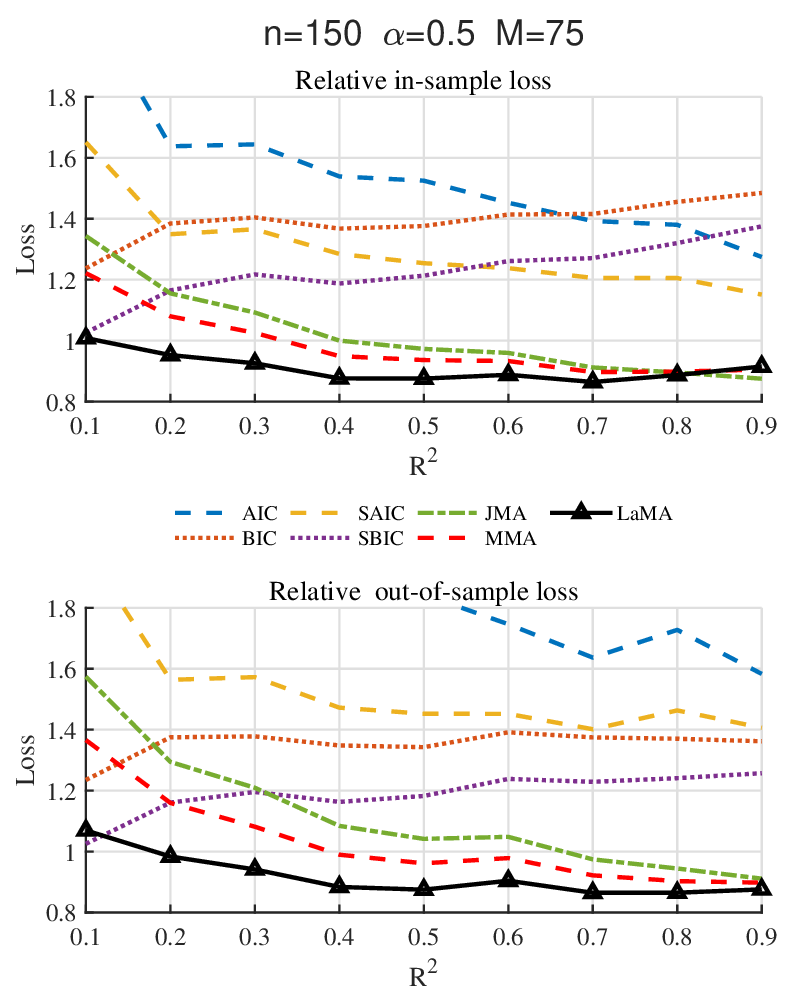}}
	\subfloat{\includegraphics[scale=0.35]{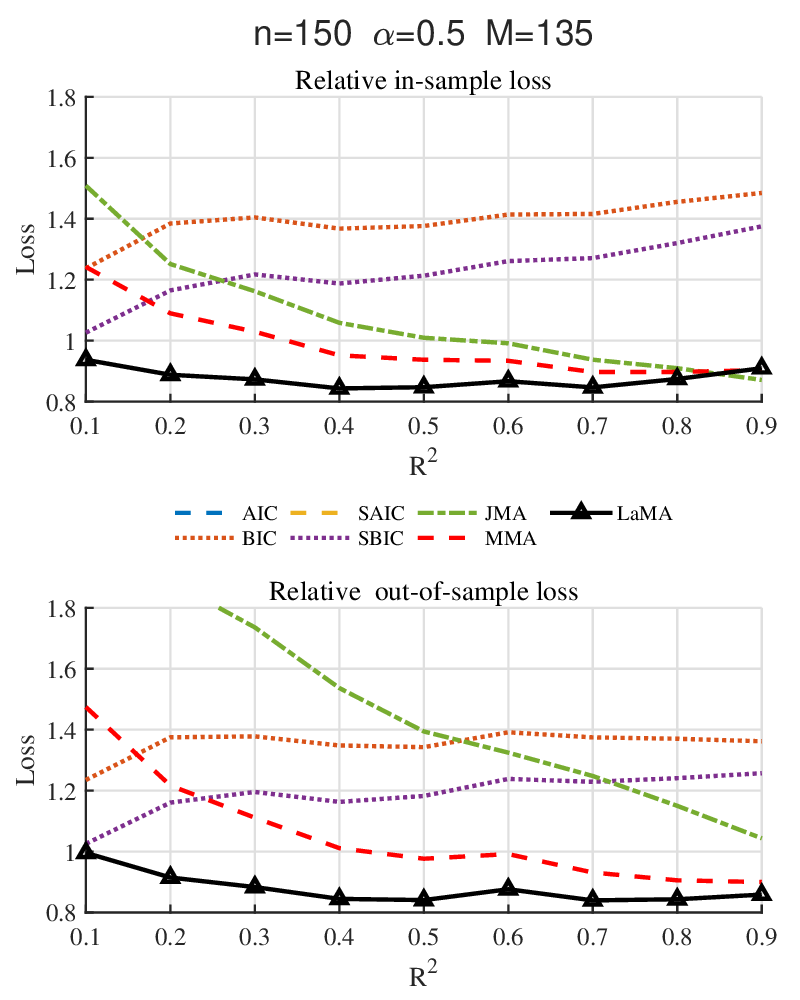}}
	\setlength{\abovecaptionskip}{0pt}
	\caption{ $n=150, \alpha=0.5, M\in \{ 16, 75, 135\}.$}
	\label{fig: n150}
\end{figure*}
\begin{figure*}[htb]
	\centering
	\subfloat{\includegraphics[scale=0.35]{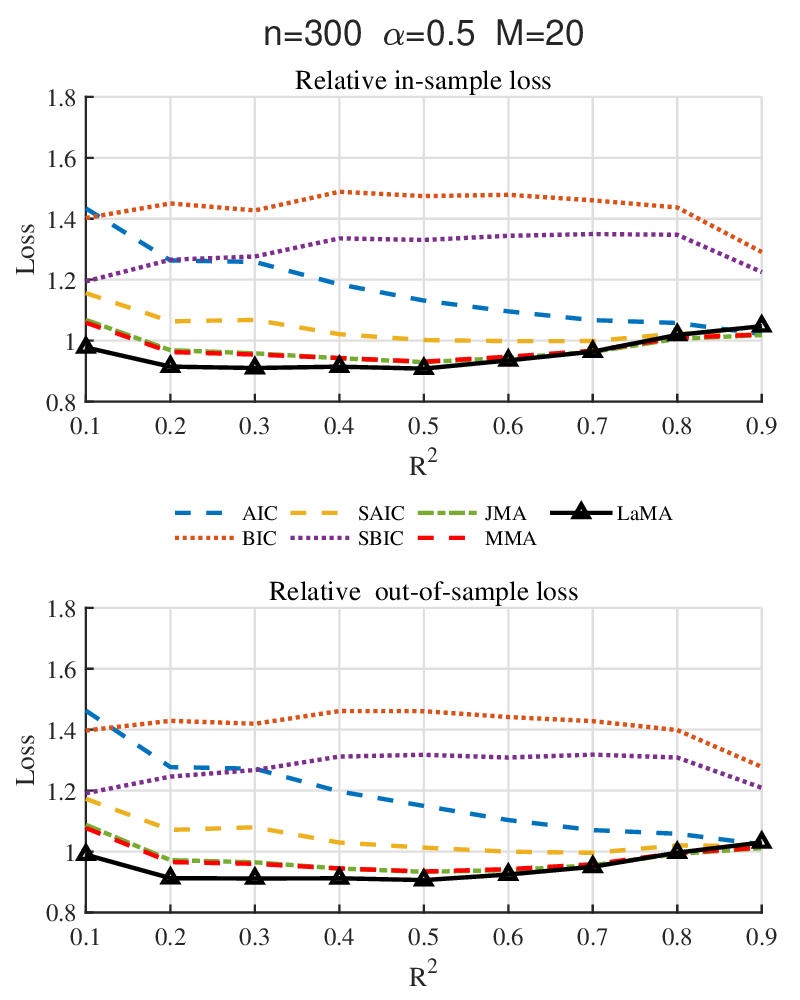}}
	\subfloat{\includegraphics[scale=0.35]{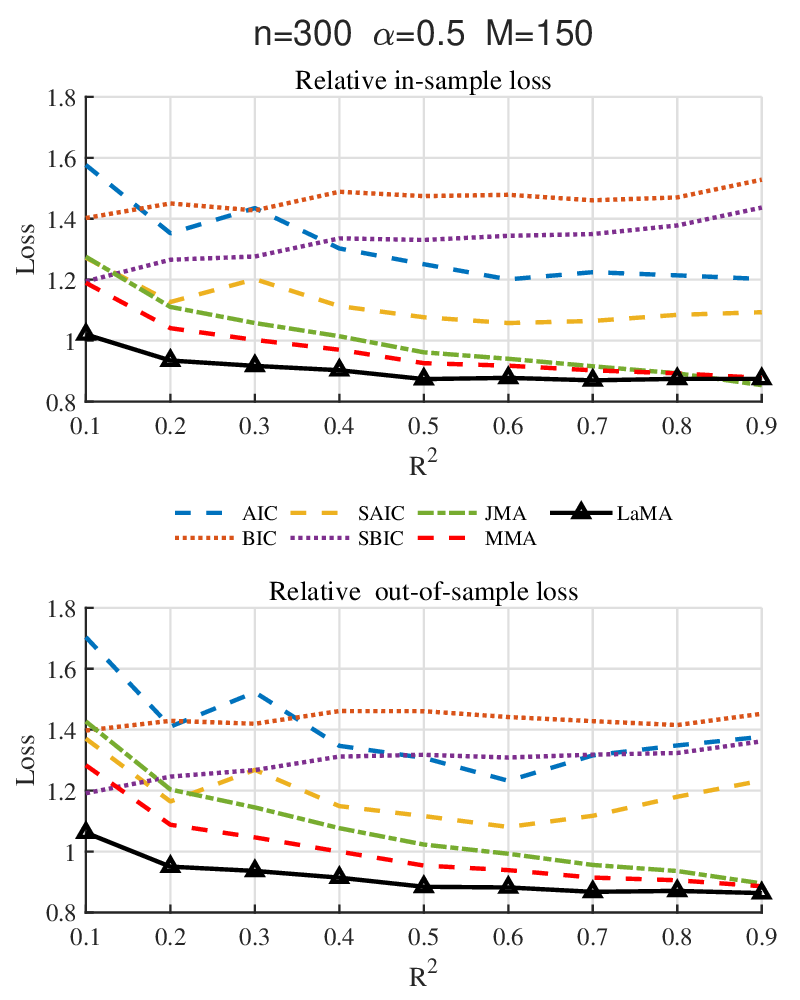}}
	\subfloat{\includegraphics[scale=0.35]{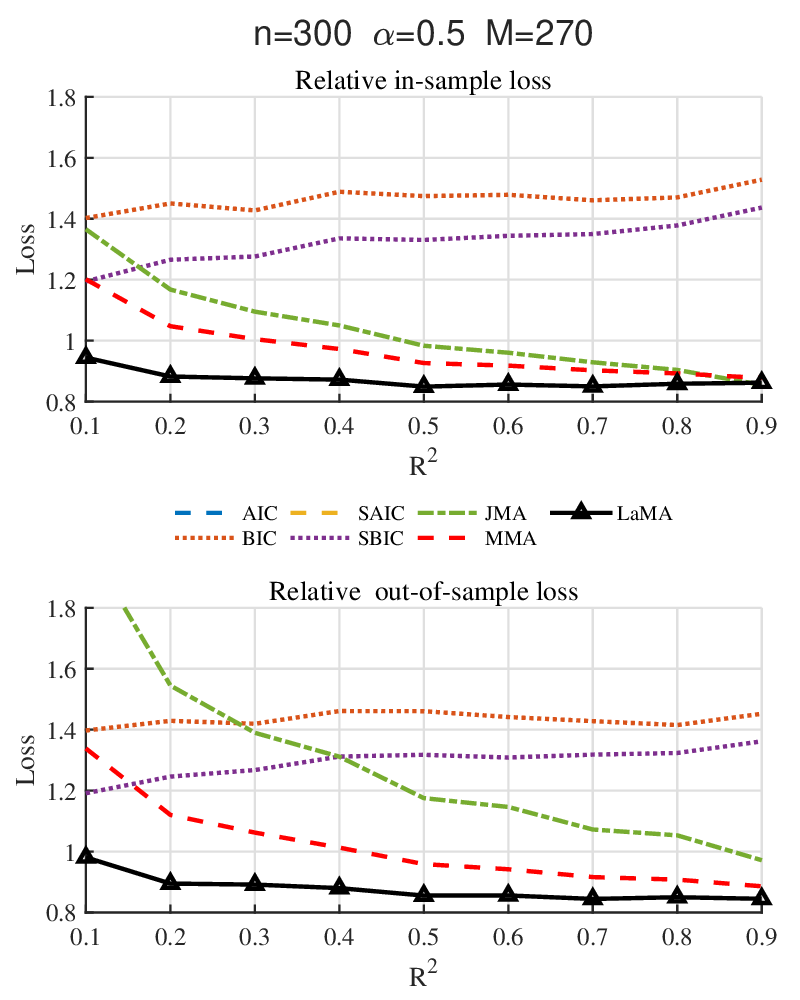}}
	\setlength{\abovecaptionskip}{0pt}
	\caption{ $n=300, \alpha=0.5,M\in \{ 20,150, 270\}.$}
	\label{fig: n300}
\end{figure*}

From Figures \ref{fig: n150} and \ref{fig: n300}, it can be observed that as the sample size increases to $n = 150$ and $n=300$, the performance gap among different methods narrows progressively. 
This confirms that as the sample size grows, asymptotic theory becomes increasingly relevant, and competing methods begin to recover their theoretical properties. 
Nevertheless, LaMA continues to maintain the lowest out-of-sample loss in the vast majority of settings.
While its comparative advantage is relatively less pronounced in strictly low-dimensional regimes, LaMA's performance remains outstanding as the maximum model dimension approaches the sample size. 
Specifically, MMA performs better than JMA overall, but its out-of-sample loss shows increased volatility and deviates from the optimal level as dimensionality increases. 
Furthermore, the experimental results characterize the sensitivity of these methods to SNR. As $R^2$ decreases, which corresponds to a weaker signal given a fixed noise variance, the loss curve of LaMA remains noticeably flatter compared to alternative approaches. 
This indicates that our method is less sensitive to high-noise conditions under a large ratio of $M/n$.
The underlying reason for this phenomenon lies in the design of LaMA: it explicitly accounts for both in-sample bias and out-of-sample variance when determining model weights. 
Its adaptive regularization imposes a heavier penalty on candidate models with high variance, effectively suppressing the variance inflation commonly encountered in high-dimensional environments. 
In contrast, MMA minimizes an estimate of in-sample risk, which leads to an underestimation of out-of-sample variance as the number of models grows. 
Similarly, while JMA approximates test-set information through a jackknife procedure, the lack of an explicit variance-stabilizing constraint renders it more susceptible to noise relative to LaMA in high-dimensional settings.

\section{Empirical analysis}
In this section, the proposed LaMA method is applied to two classic datasets available in R: the U.S. Crime dataset (47 observations, 15 features) and the Motor Trend Car dataset (32 observations, 10 features). 
The regressors are prioritized through Mallows' $C_p$-based forward selection, iteratively selecting those that minimize $C_p$, thereby constructing a deterministic sequence of nested candidate models.
For each dataset with total sample size $N$, we randomly partition the observations into a training set of size $n$ and a test set of size $N-n$. 
To eliminate the influence of random sample splitting, this process is repeated 10,000 times. 
We evaluate the prediction generalization ability and stability using the mean and variance of the empirical test error: 
$ \|\hat{\bm{\mu}}_{\text{test}}(\bm{\omega}) - \bm{Y}_{\text{test}}\|^2/(N-n),$
where $\bm{Y}_{\text{test}}$ denotes the observed responses in the test set.

\begin{table}[htb]
	\caption{Test error on the U.S. Crime dataset.}	\label{tab: US}
	\renewcommand\arraystretch{1}
	{	\begin{threeparttable}
			\begin{tabular}{ccccccccc}
				\hline
				& $n$  &  AIC  &  BIC  &  SAIC & SBIC   & MMA &  JMA &   LaMA   \\
				\hline
				Mean     &    18 & 5.1297 & 4.3753 & 4.6798 & 3.8803 & 2.1516 & 0.9997 & \textbf{0.6777} \\
				& 21 & 1.4712 & 1.0971 & 1.2861 & 0.9412 & 0.7742 & 0.6519 & \textbf{0.5526} \\
				& 24 & 0.9237 & 0.7039 & 0.8073 & 0.6127 & 0.5830 & 0.5601 & \textbf{0.4983} \\
				& 27 & 0.6963 & 0.5735 & 0.6197 & 0.5096 & 0.5079 & 0.5075 & \textbf{0.4620} \\
				& 	30 & 0.5736 & 0.5114 & 0.5217 & 0.4589 & 0.4652 & 0.4701 & \textbf{0.4347} \\
				& 33 & 0.5140 & 0.4788 & 0.4742 & 0.4346 & 0.4428 & 0.4482 & \textbf{0.4175} \\
				\hline
				Variance&	18 & 267.6944 & 243.1452 & 240.5451 & 207.0834 & 74.7662 & 7.6021 & \textbf{0.5919} \\
				&21 & 2.7682 & 1.7666 & 2.0920 & 1.3189 & 0.5988 & 0.2863 & \textbf{0.1721} \\
				&24 & 0.6714 & 0.3668 & 0.4707 & 0.2620 & 0.1802 & 0.1403 & \textbf{0.1042} \\
				&	27 & 0.2620 & 0.1315 & 0.1905 & 0.1000 & 0.0904 & 0.0862 & \textbf{0.0720} \\
				&30 & 0.1434 & 0.0769 & 0.1072 & 0.0629 & 0.0654 & 0.0699 & \textbf{0.0610} \\
				&33 & 0.0939 & 0.0591 & 0.0747 & \textbf{0.0524} & 0.0540 & 0.0600 & 0.0526 \\
				\hline
			\end{tabular}
			\begin{tablenotes}
				\item For each training set size $n$, the minimum values of the test error mean and variance are highlighted in bold.
			\end{tablenotes}
	\end{threeparttable} }
\end{table}

\begin{table}[htb]
	\caption{Test error on the Motor Trend Car dataset.}	\label{tab: Car}
	\resizebox{\linewidth}{!}{	\renewcommand\arraystretch{1}
		{
			\begin{threeparttable}
				\begin{tabular}{ccccccccc}
					\hline
					& $n$  &  AIC  &  BIC  &  SAIC & SBIC   & MMA &  JMA &   LaMA \\
					\hline
					Mean    &   13 & 5.6253 & 5.0062 & 4.8457 & 4.2394 & 2.1088 & 0.5924 & \textbf{0.4193} \\
					&16 & 1.0020 & 0.7469 & 0.8002 & 0.6019 & 0.4983 & 0.3854 & \textbf{0.3075} \\
					&19 & 0.5090 & 0.3623 & 0.4135 & 0.3193 & 0.3182 & 0.3108 & \textbf{0.2674} \\
					&	22 & 0.3662 & 0.2715 & 0.3056 & 0.2537 & 0.2676 & 0.2752 & \textbf{0.2452} \\
					&	25 & 0.3052 & 0.2391 & 0.2584 & \textbf{0.2287} & 0.2444 & 0.2498 & 0.2291 \\
					&	28 & 0.2820 & 0.2290 & 0.2404 & \textbf{0.2203} & 0.2375 & 0.2392 & 0.2237 \\
					\hline
					Variance     & 	 13 & 3381.7649 & 3341.1992 & 3185.8266 & 2877.5550 & 350.3377 & 2.8595 & \textbf{0.4825} \\
					&16 & 4.0567 & 3.0680 & 2.5043 & 1.6448 & 0.7243 & 0.1904 & \textbf{0.0618} \\
					&19 & 0.5232 & 0.2945 & 0.3233 & 0.1913 & 0.1147 & 0.0671 & \textbf{0.0365} \\
					&22 & 0.1356 & 0.0519 & 0.0698 & 0.0361 & 0.0337 & 0.0347 & \textbf{0.0262} \\
					&	25 & 0.0705 & 0.0280 & 0.0335 & \textbf{0.0232} & 0.0246 & 0.0266 & 0.0243 \\
					&	28 & 0.0534 & 0.0352 & 0.0344 & \textbf{0.0316} & 0.0331 & 0.0340 & 0.0338 \\
					\hline
				\end{tabular}
				\begin{tablenotes}
					\item For each training set size $n$, the minimum values of the test error mean and variance are highlighted in bold.
				\end{tablenotes}
	\end{threeparttable} }}
\end{table}

Table \ref{tab: US} and Table \ref{tab: Car} summarize the empirical test error results across various training sample sizes for both datasets. 
When the training sample size $n$ is small and comparable to the total number of features, 
LaMA exhibits an advantage with  lower error means and variances than competing methods. 
Other methods exhibit severe instability, characterized by inflated variances, which stems from the variance explosion in out-of-sample risk. 
This indicates that LaMA effectively extracts information and maintains high stability even in the presence of candidate models near the interpolation boundary.
As the training sample size $n$ increases, the regression moves into a more traditional under-parameterized regime. 
Consequently, the performance gaps among the evaluated methods gradually narrow, with SBIC yielding slightly lower errors than LaMA in larger-sample settings. 
Nevertheless, LaMA remains highly competitive and exhibits comparable performance in these scenarios, offering early stability even without a large sample size.

\section{Conclusion and discussion}
This paper provides a rigorous asymptotic characterization of out-of-sample risk in high-dimensional model averaging, by leveraging  random matrix limit theory. 
Our analysis reveals that the ensemble using a simple weighting strategy exhibits a double-descent phenomenon, characterized by variance explosions near the interpolation boundary. 
In contrast, model averaging that allocates weights  based on the ``high-risk, low-weight'' principle can completely suppress this risk peak, thereby transforming the landscape into a well-behaved, globally flat surface.
We term this phenomenon ensemble emergence.
Building on these insights, we propose LaMA for settings where the dimension of the largest model, $k_M$, and the sample size, $n$, are comparable.
Moving beyond the classical $k_M \ll n$ constraint, LaMA replaces conventional in-sample risk minimization with a regularized trade-off between estimation of in-sample bias and asymptotic out-of-sample variance. 
By incorporating this explicit variance penalty, LaMA provides a theoretically grounded framework that maintains generalization capabilities even in severely high-dimensional regimes where traditional methods typically diverge.

A central discussion in model averaging concerns asymptotic optimality. 
Traditional frameworks \cite{MMA2007, WAN2010277} have historically relied on technical conditions that restrict the growth of model complexity relative to the sample size.
Even as recent extension \cite{RePEc2021} allows for a larger number of regressors, there are still restrictions on the total number of candidate models.
In the context of nested models, where the number of models $M$ is directly proportional to the dimension of the largest model $k_M$, these classical assumptions are violated when the ratio $k_M/n \to c_M \in (0, 1)$.
Consequently, traditional optimality guarantees lose their validity in such high-dimensional regimes.
A recent significant work by \cite{Peng03042025} aims to relax these constraints, suggesting that previous restrictions may have obscured the full oracle potential of model averaging.
In particular, their theory no longer requires a direct upper bound on the number of candidate models \(M\). 
Instead, the validity of asymptotic optimality is ensured through more intrinsic conditions, including appropriate control of the complexity of the nested candidate path, sufficient accuracy of the variance estimator, and suitable regularity assumptions on the signal sequence. 
This development substantially relaxes the classical restrictions on candidate-model growth and shows that, at least for in-sample risk, model averaging can remain theoretically optimal even when the largest nested model is comparable to the sample size.

Despite these advancements, existing optimality theories remain primarily confined to in-sample risk.
As this study highlights, the trajectories of in-sample and out-of-sample risks  diverge sharply in high-dimensional settings. 
An estimator that achieves asymptotic unbiasedness for in-sample risk often suffers from uncontrolled variance explosion in its out-of-sample predictions. 
Recognizing this discrepancy, the design of LaMA deliberately shifts from the rigid pursuit of unbiasedness toward stringent control over variance peaks. 
While we demonstrate the practical predictive efficacy of this approach, establishing a formal proof of asymptotic optimality specifically for out-of-sample risk within the $k_M/n \to c_M$ regime remains a significant open challenge.

\section*{Acknowledgments}
	The authors would like to thank the Editor, the Associate Editors, and the referees for their review of the paper.

\section*{Funding}
	The authors were supported by NSFC under Grant Nos. 72525001, 72495124, and 12571311.

\section*{Supplementary material}\label{supp}
The supplementary material includes the detailed decomposition of out-of-sample risk, the proofs of Theorems \ref{Th: out-of-sample risk} and \ref{Th: trace}, the experiments on randomly weighted model averaging, the algorithm for LaMA method, the calculation of in-sample risk, and additional numerical results under the alternative variance estimator.

\bibliographystyle{elsarticle-num} 
\bibliography{bibliography} 


\end{document}